\documentclass[twocolumn,aps,prd,preprintnumbers,bibnotes11pt,superscriptaddress,amsmath,amssymb,nofootinbib]{revtex4-1}

\usepackage{graphicx,subfigure}
\usepackage{float}
\usepackage{color}
\usepackage{epstopdf}
\usepackage{amsmath}
\usepackage{amsthm}
\usepackage{amssymb}
\usepackage{amsfonts}
\usepackage{graphicx}
\usepackage{txfonts}
\usepackage{ulem}
\usepackage{mathtools}
\usepackage{bm}

\usepackage{dcolumn}

\usepackage[colorlinks=true,urlcolor=blue,citecolor=blue,linkcolor=red]{hyperref}
\usepackage{booktabs}
\usepackage{braket}
\usepackage[dvipsnames]{xcolor}
\usepackage{ulem}

\usepackage{amsmath}
\usepackage{graphicx}
\usepackage{ulem}
\usepackage{dcolumn}
\usepackage{epsfig}
\usepackage{bm}
\usepackage{array}
\usepackage{hyperref}
\hypersetup{
colorlinks=true,
citecolor=blue,
linkcolor=blue,
urlcolor=black,
pdfmenubar=true
}

\usepackage{hyperref}
\usepackage{graphicx}
\usepackage{amsmath}
\usepackage{amsfonts}
\usepackage{amssymb}
\usepackage{xcolor}
\usepackage{bbm}
\usepackage{calrsfs}
\usepackage{dutchcal}
\usepackage{amsthm}

\usepackage{tikz}
\usetikzlibrary {arrows.meta}

\newcommand{\be}{\begin{equation}}
\newcommand{\ee}{\end{equation}}
\newcommand{\ii}{\text{i}}
\newcommand{\dd}{\text{d}}

\newcommand{\Line}[2]{\raisebox{2pt}{\tikz{\draw[-,#1,#2,line width=1pt](0,0) -- (3mm,0);}}}
\newcommand{\Arrow}[2]{\tikz{\draw[arrows={-Latex[width=6pt, length=4pt]},#1,#2,line width=1pt](0,0) -- (2.5mm,0);\draw[-,#1,#2,line width=1pt](1mm,0) -- (3mm,0);}}

\definecolor{blue}{HTML}{1F77B4}
\definecolor{orange}{HTML}{FF7F0E}
\definecolor{green}{HTML}{2CA02C}
\definecolor{red}{HTML}{D62728}
\definecolor{purple}{HTML}{9467BD}
\definecolor{gray}{HTML}{7F7F7F}

\usepackage{changepage}

\begin{document}

\title{Platform and environment requirements of a\\ satellite quantum test of the Weak Equivalence Principle at the $10^{-17}$ level}

\author{Christian Struckmann$^\dagger$}\affiliation{Leibniz University Hannover, Institute of Quantum Optics, Welfengarten 1, 30167 Hannover, Germany}
\author{Robin Corgier$^\dagger$}\affiliation{LNE-SYRTE, Observatoire de Paris, Universit\'e PSL, CNRS, Sorbonne Universit\'e 61 avenue de l'Observatoire, 75014 Paris, France}
\author{Sina Loriani}\affiliation{Potsdam Institute for Climate Impact Research, Member of the Leibniz Association, Telegrafenberg 31A, 14473 Potsdam}
\author{Gina Kleinsteinberg}\affiliation{Leibniz University Hannover, Institute of Quantum Optics, Welfengarten 1, 30167 Hannover, Germany}
\author{Nina Gox }\affiliation{Leibniz University Hannover, Institute of Quantum Optics, Welfengarten 1, 30167 Hannover, Germany}
\author{Enno Giese}\affiliation{Technische Universit{\"a}t Darmstadt, Fachbereich Physik, Institut f{\"u}r Angewandte Physik, Schlossgartenstr. 7, D-64289 Darmstadt, Germany}
\affiliation{Leibniz University Hannover, Institute of Quantum Optics, Welfengarten 1, 30167 Hannover, Germany}
\author{Gilles Métris}\affiliation{Universit\'e C\^ote d'Azur, Observatoire de la C\^ote d'Azur, CNRS, IRD, G\'eoazur, 250 avenue Albert Einstein, F-06560 Valbonne, France}
\author{Naceur Gaaloul}\affiliation{Leibniz University Hannover, Institute of Quantum Optics, Welfengarten 1, 30167 Hannover, Germany}
\author{Peter Wolf}\affiliation{LNE-SYRTE, Observatoire de Paris, Universit\'e PSL, CNRS, Sorbonne Universit\'e 61 avenue de l'Observatoire, 75014 Paris, France}

\begin{abstract}
    The Space Time Explorer and QUantum Equivalence principle Space Test (STE-QUEST) recently proposed, aims at performing a precision test of the weak equivalence principle (WEP), a fundamental cornerstone of General Relativity. Taking advantage of the ideal operation conditions for high-precision quantum sensing on board of a satellite, it aims to detect possible violations of WEP down to the $10^{-17}$ level.
    This level of performance leads to stringent environmental requirements on the control of the spacecraft. We assume an operation of a dual-species atom interferometer of rubidium and potassium isotopes in a double-diffraction configuration and derive the constraints to achieve an Eötvös parameter $\eta=10^{-17}$ in statistical and systematic uncertainties. We show that technical heritage of previous satellite missions, such as MICROSCOPE, satisfies the platform requirements to achieve the proposed objectives underlying the technical readiness of the STE-QUEST mission proposal.
\end{abstract}

\maketitle
\def\thefootnote{$\dagger$}\footnotetext{These authors contributed equally.}

\section{Introduction}

The fundamental physics of nature is described by General Relativity (GR) and the Standard Model of particle physics (SM)~\cite{Kiefer,Will2018}. 
Both theories have been separately extensively tested without showing any discrepancy but their unification  remains an unresolved problem.
The validity of GR at the quantum level is still unknown and the discovery of new forces beyond the SM is not excluded. 
Moreover, the SM accounts only for the visible matter in the Universe, while the dominant component of matter is dark and its quantum nature is still unclear. 
On the other hand, the SM and quantum mechanics are very successful at explaining the microscopic phenomena, but also pose fundamental questions such as the measurement problem and the quantum-classical transition. 
The ultimate theoretical challenge may be to construct a theory of quantum gravity that reconciles SM and GR, which may require modifying or extending one or both of these frameworks.
Several quantum gravity models, unifying all non-gravitational interactions with gravity predict a violation of the Einstein Equivalence Principle (EEP), a cornerstone of GR, yet not a fundamental symmetry of Nature. 
It is consequently of fundamental importance to search for possible violations of the EEP, which has three facets~\cite{10.1119/1.4895342}: Local Lorentz Invariance, Local Position Invariance and Universality of Free Fall, also referred to as Weak Equivalence Principle (WEP). 
Schiff's conjecture speculates that a violation of one implies the violation of the two others \cite{Will2018}.
If the WEP holds, the trajectory of a freely falling, uncharged test-body only depends on the initial position and velocity of the test-body but is independent of its mass, composition, form or spin \cite{Will2018}. 
A convenient figure of merit for all WEP tests is the E\"otv\"os ratio $\eta$. It quantifies the differential free-fall acceleration of two test masses of different composition, thereby measuring a possible violation of the WEP. 
Although it is a useful tool for comparing different experiments, it cannot account for the diversity of possible underlying theories, e.g., different types of couplings depending on the source and test objects, or couplings to space-time-varying background fields other than local gravity.
Thus, not only the best performance in the E\"otv\"os ratio is required, but also a large diversity of test objects and source masses of different nature.

At what performance of a WEP test do we expect to see a violation? There is no firm and widely accepted value, but a number of models predict violations in the $10^{-10}  - 10^{-22}$ region based on unification scenarios \cite{damour:1994fk}, supersymmetry and dark matter \cite{fayet:1974ww,fayet:1990tu,Fayet:2017pdp}, or Lorentz symmetry breaking at the Planck scale \cite{Kostelecky:1994rn}. If one also takes into account cosmological inflation scenarios the possible region of WEP breaking is reduced to $10^{-10}  - 10^{-19}$ \cite{damour:2002ys,damour:2010zr,Fayet:2017pdp,fayet:2019aa}.

Much of this region (down to $10^{-15}$) is already excluded by experiments. A major discovery may thus be ``just around the corner''. Today the best result with classical test masses has been obtained with the MICROSCOPE space mission at the level of $\eta = [-1.5 \pm 2.3(stat) \pm 1.5(syst)] \times 10^{-15}$~\cite{Touboul2022} while equivalent ground tests are ultimately limited by the Earth’s gravitational environment to $\eta\approx10^{-13}$~\cite{Wagner2012,Williams09}. 

Quantum tests of the WEP (Q-WEP) can be performed through matter wave interferometry where precision measurements are obtained by mapping the physical quantity of interest (the acceleration) to a phase shift determined using interferometric techniques.  
Matter-wave interferometers played a key role in the development of quantum theory~\cite{jonsson1961elektroneninterferenzen,gerlach1922experimentelle} and have been widely used to accurately determine the fine structure constant~\cite{Parker18, Morel20} or the gravitational constant~\cite{Sorrentino14, Rosi14}.
Besides testing fundamental laws of nature, quantum sensors have been developed to measure inertial forces and used as gravimeters, gradiometers, and gyroscopes~\cite{Geiger20}.
Furthermore, in order to exploit the enhanced sensitivity of long interrogation times, large-scale setups are currently in planning or construction for the detection of gravitational waves and dark matter \cite{Canuel2017,Canuel20,Badurina20,Abe2021,Zhan2020,ArxivCERN}.
Indeed, long free fall interrogation times already enable a Q-WEP on ground at the level of $\eta = [1.6 \pm 1.8(stat) \pm 3.4(syst)] \times 10^{-12}$ with different atomic isotopes~\cite{PhysRevLett.112.203002,PhysRevLett.115.013004,Asenbaum2020,Zhang_2020,10.1116/5.0076502}, made possible by the extremely low expansion energies accessible with ultracold ensembles \cite{Kovachy2016, Corgier18, Corgier20, Deppner2021, Gaaloul22}. 
Longer interrogation times of some tens of seconds being accessible in space, unlock the potential of Q-WEP tests at the level of $\eta$ in the range of $10^{-15}$ to $10^{-17}$, as explored in this paper.
This outlook is supported by the significant progress made in the last decade on the technological readiness level of cold and ultra-cold atomic inertial sensors, as demonstrated in micro-gravity experiments in 0-g flights~\cite{Condon2019,barrett2016,Geiger2011}, drop-towers~\cite{Deppner2021,Rudolph2015,vanZoest2010}, sounding rockets~\cite{Vogt2020, Kulas2017, becker2018, Lachmann2021} and  on-board the International Space Station~\cite{aveline2020,Gaaloul22,JPL23}.

In this article, we investigate the realistic requirements on the atom interferometer and the spacecraft platform to perform a space-borne Q-WEP test with a dual-species atom interferometer of ${}^{87}$Rb and ${}^{41}$K isotopes. 
In Sec.~\ref{sec:inertial-sensor} we describe the working principle of the inertial sensor. In Sec.~\ref{sec:environment-constraints} we investigate the constraints on the atom interferometer environment and those on the satellite control in Sec.~\ref{sec:spacecraft-constraints}. 
A discussion about the feasibility of a Q-WEP satellite-borne test is discussed in Sec.~\ref{sec:feasibility} before concluding in Sec.~\ref{sec:conclusion}.

\section{Atom Interferometry}\label{sec:inertial-sensor}

\subsection{Principle}
\begin{figure}[h!]
\includegraphics[width=\linewidth]{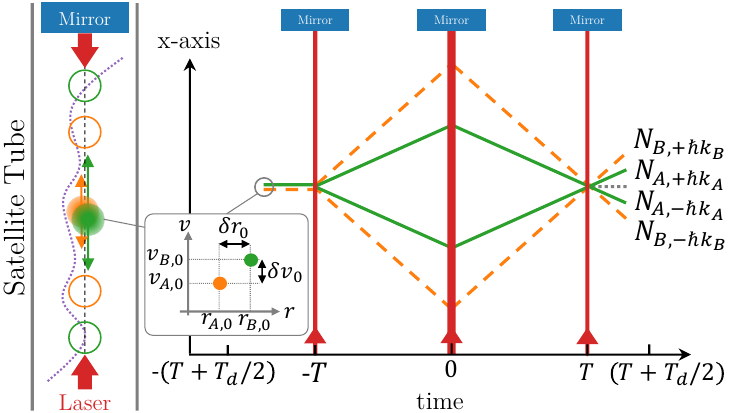}
\caption{Dual-species atom interferometer sequence. The space-time diagram shows the trajectories of the two species $A\equiv \mathrm{rubidium}$ (\protect\Line{green}{solid}) and $B\equiv \mathrm{potassium}$ (\protect\Line{orange}{dashed}) atoms. The atomic ensemble of each species are split in a superposition of momentum state, redirected and recombined using double diffraction $\pi/2$, $\pi$ and $\pi/2$ pulses at times $[-T,0,T]$ (\protect\Arrow{red}{solid}). Note that the pulse separation time $T$ can differ between both species, but we choose $T_A=T_B=T$ (see Sec.~\ref{sec:acceleration}). $T_\mathrm{d}$ denotes the dead-time and is equally split between the state preparation and the detection time as an illustration. The output ports of each species interferometer can be distinguished by the different external states $\pm \hbar k_i$ and $0\hbar k_i$. The presence of an external potential is highlighted by the dotted line in the satellite tube along the sensitive axis of the interferometer (\protect\Line{purple}{densely dotted}).}
\label{fig:ati-sequence}
\end{figure}

Dual-species atom interferometers are powerful tools for measuring differential acceleration by exploiting quantum-mechanical effects.
In this study, we focus on a Mach-Zehnder double-diffraction configuration~\cite{levequePRL,AhlersPRL,Hartmann2020,PhysRevLett.115.013004}.  
The interferometric sequence of each species, denoted by $A$ and $B$, consists of three atom-light interaction pulses as illustrated in Fig.~\ref{fig:ati-sequence}. 
A first $\pi/2$-pulse creates a coherent quantum superposition of momentum states and leads to a spatial separation of the interferometer arms. 
The trajectories are then reflected by a $\pi$-pulse and finally recombined at a final $\pi/2$-pulse. 
In between the pulses, each superposition freely evolves for a duration $T_i$ and accumulates a phase leading to a final phase difference $\Phi_i$, $i$ being $A$ or $B$. 
That phase difference is evaluated by measuring the relative atom numbers at the output ports of each interferometer, which differ only in momentum in the case of Bragg atom-light diffraction or in momentum and internal states in the case of Raman diffraction~\cite{Hartmann2020}.
We write the general linear phase combination, $\Phi_{\rm gen}$, as

\begin{equation}
\label{phi_gen}
    \Phi_{\rm gen} = \mathcal{A} \Phi_{\rm A} + \mathcal{B} \Phi_{\rm B}.
\end{equation}

Here $\mathcal{A}, \mathcal{B}$ are freely selectable constants in the data analysis and $\Phi_{\rm i}$ are the unmodified raw data of the interferometer instrument.
Note that the interrogation time of each species can differ, $T_A\neq T_B$.

The atom interferometer phase is related to the acceleration in the local inertial frame through the relation, 
\begin{equation}
    \Phi_i = 2 k_i a_i T_i^2,
    \label{eq_AI_phase}
\end{equation}

where $k_i=4\pi/\lambda_i$ is the effective wave number and $a_i$ is the total acceleration experienced by each species.
We further decompose the acceleration $a_i$, as the sum of the gravitational acceleration, $g_i=g_0+\gamma_i$, where $g_0$ is the universal gravitational acceleration at the location of the experiment (see Tab.~\ref{tab:params}) and $\gamma_i=\mathcal{g}_i+a_{b,i}$ is the deviation of species $i$. This deviation encompasses the hypothetical Q-WEP violation we want to extract, $\mathcal{g}_{i}$, plus all spurious bias phase shift terms $\Phi_{b,i}$ that can be interpreted as an acceleration of the form $a_{b,i}=\Phi_{b,i}/2k_i T_i^2$.
The sensitivity on the Eötvös parameter, $\eta$,
\begin{equation}
\label{eq_eotvos}
    \eta =\dfrac{\gamma_A-\gamma_B}{g_0},
\end{equation}
is bound either by the uncertainty on the bias acceleration $\Delta\delta a_b$, i.e., the uncertainty of $\delta a_b=a_{b,A}-a_{b,B}$\footnote{Throughout this paper we will use $\delta$ to represent the difference of two quantities, and $\Delta$ to indicate the uncertainty of a quantity.}, or by the uncertainty on the measurement itself, limited by the standard quantum limit for classically correlated particles.
This paper aims to study the different contributions in $a_{b,i}$ coming from the interferometer environment and/or the satellite platform to highlight the requirements to test the Q-WEP at different values of the Eötvös parameter.

\subsection{Signal demodulation in satellite setups}\label{sec:Adv_Sat}
Space-borne platforms allow for a long interrogation time, $T_i\gg 1$\,s, and therefore high sensitivities (see Eq.~\eqref{eq_AI_phase}). 
Additionally, for space-borne setups, the projection of the gravitational potential onto the sensitive measurement axis depends on the position and attitude of the satellite. As a consequence, the differential phase shift $2 \eta g_0 k T^2$ is naturally modulated at certain frequencies.
For example, in the case of circular orbit with inertial attitude, $g(t)$ is modulated at orbital frequency, $f_{\rm orb} = \omega_{\rm orb} / 2\pi$. 
Systematic effects modulated at different frequencies can therefore be reduced by at least $2/(\mathcal{N}_c T_c\omega_{\rm orb})$ where $\mathcal{N}_c$ denotes the total number of measurements and $T_c$ the cycle time of the measurement sequence~\cite{loriani2020, Ahlers2022}.
Thus, the more stringent requirements are only on systematic effects modulated at $f_{\rm orb}$. 

In more detail, several strategies, listed below, can be employed to drastically reduce the impact of a parasitic systematic effect on the desired signal. 
It is worth to note that these strategies can potentially accumulated. 
Here, we consider a parasitic effect at frequency $f_p$ and amplitude $A_p$, searching for a signal at frequency $f_{sig}$:

(i) If $f_{p}$ differs from $f_{sig}$, the perturbation can be de-correlated from our science signal provided that $|f_p-f_{sig}| > 1/T_{sc}$, where $T_{sc}$ denotes the science time. 
Consequently, for any parasitic periodic effect of amplitude $A_{p}$ at frequency $f_{p}$, one only needs to consider its amplitude $A_{sys}$ at $f_{sig}$.
For example, as observed in the MICROSCOPE mission~\cite{Touboul2017, Touboul2022}, the DC self-gravity perturbation and its residual effect at $f_{sig}$, based on a typical thermal expansion coefficient of $10^{-5}$/K for the satellite, and a typical peak to peak temperature variation of about $1$~K at orbital frequency lead to a reduction factor of about $10^5$.
A more precise evaluation would require a detailed design and thermoelastic model of the satellite, and is beyond the scope of this paper. 

(ii) The effect of $A_{sys}$ can further be reduced by a likely phase mismatch and controlled phase jumps, which will be present in the searched signal but unlikely to be fully present in the systematic effect. 
An interesting strategy, which consists of rotating the satellite by a fixed angle $\theta_R$ every given $N_{\rm orb}$ orbits, is discussed in more detail in Sec.~\ref{sec_attitude_control}. 
Such a strategy would relax the constraints on systematic effects modulated at $\omega_{\rm orb}$ provided they are not at all, or only partly, affected by these controlled rotations.
In Ref.~\cite{Ahlers2022} the authors estimate that this procedure could lead to a further reduction factor of about $10^{3}$, mainly limited by the imperfect knowledge of the parameters (angles, timing, \dots) and correlations of the perturbation and the induced angular steps.

(iii) Finally, if the systematic effect can be modelled, possibly with unknown parameters, its impact at $f_{sig}$ can be efficiently corrected provided this effect has also significant components at other frequencies different from $f_{sig}$, allowing the fitting of the model parameters to the data. 
A prime example of this is the effect of gravity gradients in MICROSCOPE whose amplitude $A_{sys}$ at $f_{sig}$ could be reduced by more than $10^{7}$ with respect to $A_{p}$ (at $f_p=2f_{sig}$) by fitting the model parameters to the data \cite{Touboul2017}, although the actual (unfitted) component $A_{sys}$ was only $\sim 10^{3}$ times smaller than $A_{p}$. 

To conclude, the stringent requirements that are derived in the rest of the paper will in practice be relaxed by large amounts.
The order of magnitude of those reduction factors is indicated in the individual sections. However, an exact evaluation is beyond the scope of this paper as it requires a detailed and specific satellite design.

\subsection{Parameters of the interferometer sequence}
Throughout this paper we consider a Q-WEP test at the level of $\eta=10^{-15}$ and $\eta=10^{-17}$. 
The typical mission parameters, as envisioned for the STE-QUEST space mission scenarios~\cite{loriani2020, Voyage2050,Ahlers2022}, are explicitly given in Tab.~\ref{tab:params} and divided in three categories. 
The first one refer to the satellite plate-form. 
The second one refer to interferometer sequence and include details on the atomic species. 
The last one highlight the constrains on the quantum-state engineering of the two test masses.

\begin{table}[h!]
\caption{Operational parameters of the atom interferometer to test the Q-WEP at the level of $\eta=10^{-15}$ and $10^{-17}$.}
\begin{tabular}{c|c|c}
    \hline \\ [-1.5ex]
    Parameters & $\eta=10^{-15}$ & $\eta=10^{-17}$ \\
    \hline
    \hline
    \multicolumn{3}{c}{\bf Mission}\\
    \hline
    Orbit, altitude (km) &\multicolumn{2}{c}{Circular, 1400} \\ 
    Attitude &\multicolumn{2}{c}{Inertial + modulation} \\
    Local gravity $g_0$ (m.s$^{-2}$) &\multicolumn{2}{c}{6.6} \\
    Gravity gradient $\partial g_0/(2\partial r)$ (s$^{-2}$) &\multicolumn{2}{c}{$8.5\times 10^{-7}$}\\
    Orbital frequency $f_{\rm orb}$ (Hz) &\multicolumn{2}{c}{$1.46\times 10^{-4}$}\\
    Mission duration $T_M$ (months)&\multicolumn{2}{c}{36} \\
    Science time $T_{sc}$ (months) &\multicolumn{2}{c}{24} \\
    \hline
    \multicolumn{3}{c}{\bf Interferometer}\\
    \hline
    Atom number $N$ & $1 \times 10^5$ & $2.5 \times 10^6$ \\
    Wave number $k_A$ for Rb (nm$^{-1}$) &\multicolumn{2}{c}{$2 \times2\pi/780$}  \\
    Wave number $k_B$ for K (nm$^{-1}$) &\multicolumn{2}{c}{$2 \times2\pi/767$} \\
    Interrogation time $2T$ (s) & 9 & 50 \\
    Max. separation Rb (m) & 0.11 & 0.59 \\
    Max. separation K (m) & 0.23 & 1.27  \\
    Cycle time $T_c$ (s) & 15 & 60 \\
    Total number of measurements $\mathcal{N}_{\rm c}$ & $2.5\times10^6$ & $7.9 \times 10^5$\\
    Contrast $C$ &\multicolumn{2}{c}{1} \\
    \hline
    \multicolumn{3}{c}{\bf Atomic source}\\
    \hline
    Diff. init. c.m. pos. $\delta x_0$ ($\mu$m) &\multicolumn{2}{c}{1} \\
    Diff. init. c.m. vel. $\delta v_0$ ($\mu$m.s$^{-1}$) & 1 & 0.1 \\
    Expansion energy (pK) & 50 & 10  \\
    Expansion velocity $\sigma_{v,Rb}$ ($\mu$m.s$^{-1}$) & 70 & 31 \\
    Expansion velocity $\sigma_{v,K}$ ($\mu$m.s$^{-1}$) & 101 & 45  \\
    Init. pos. spread $\sigma_{r,0}$ ($\mu$m) & 100 & 500 \\
    \hline
\end{tabular}
\label{tab:params}
\end{table}

\section{Constraints on the interferometer environment}\label{sec:environment-constraints}
In this section we now focus on the constraints specific to the interferometer environment in micro-gravity, even though our treatment can be generalized to ground-based environments.
In the following, we choose the coordinate system such that the sensitive axis of the interferometer is along the $x$-axis whereas the origin coincides with the initial center of mass (c.m.) positions of the atoms.

\subsection{Statistical error}\label{sec:statistical-error}

In the case of a classically correlated atomic ensemble, the phase sensitivity is ultimately limited to the quantum projection noise, where the statistical uncertainty per shot is defined as $\Delta \Phi_{i, \rm SN} = 1/(C_i\sqrt{N_i})$ with $C_i$ being the contrast and $N_i$ the atom number of interferometer $i$. 
For a dual-species atom interferometer, the standard quantum noise per measurement cycle is given by $(\Delta \delta\Phi_{\rm SN})^2= \mathcal{A}^2(\Delta\Phi_{A, \rm SN})^2+\mathcal{B}^2(\Delta\Phi_{B, \rm SN})^2$, following the notation of Eq.~\eqref{phi_gen}. 
In terms of a differential acceleration $\delta a$ this leads to the uncertainty
\begin{equation}
    (\Delta \delta a_{\rm SN})^2 = \left( \dfrac{\mathcal{A}\Delta\Phi_{A, \rm SN}}{2 k_A T_A ^2} \right)^2 + \left( \dfrac{\mathcal{B}\Delta\Phi_{B, \rm SN}}{2 k_B T_B ^2} \right)^2.
\end{equation}

Integrating the measurement over $\mathcal{N}_{\rm c}=T_{\rm sc}/T_{\rm c}$ repetitions, where $T_{\rm sc}$ is the total measurement time, $T_c = 2T+T_\mathrm{d}$ is the cycle time and $T_\mathrm{d}$ is the dead time, leads to

\begin{equation}\label{eq:eta_SN}
     (\Delta\eta)^2 = 2 \dfrac{(\Delta \delta a_{\rm SN})^2}{g^2_0\,\mathcal{N}_c(T)}.
\end{equation}
The extra coefficient 2 accounts for the sinusoidal varying local value of the gravitational acceleration due to a circular orbit~\cite{loriani2020}. Evaluating Eq.~\eqref{eq:eta_SN} with the parameters of Tab. \ref{tab:params} \footnote{We choose the {\it acceleration free combination} $\mathcal{A} = 2k_B/(k_A+k_B)$ and $\mathcal{B} = -2k_A/(k_A+k_B)$, see Sec. \ref{sec:acceleration}, thus $\mathcal{A}\approx-\mathcal{B}\approx 1$.} shows that shot noise is below the goal for $\Delta\eta$ with some margin. In continuous operation the goal is reached in $\sim 12$ months for the $\eta=10^{-15}$ case and $\sim 20$ months for the $\eta=10^{-17}$ one, well below the assumed 24 months science time. 
\subsection{Systematic effects}\label{sec:systematic-effects}
The presence of any kind of potential contributes to the interferometer phase and can lead to bias acceleration terms, ultimately limiting the sensitivity to the Eötvös coefficient of Eq.~\eqref{eq_eotvos}.
Contributions to bias phase terms are of two kinds.
On the one hand, there are effects coming from the contribution of potential gradients, acting as forces. On the other hand, there are effects coming from the presence of potential energy differences inducing Aharonov-Bohm-like phase shifts~\cite{Overstreet2022}. 
Effects of the first kind directly act on the mean trajectories of the matter-wave while the ones of the second type do not.
In this section, we analyse the bias phase terms induced by an arbitrary potential and derive constraints for specific effects. 

\subsubsection{Model of the phase accumulation}

It should be noted that different approaches have been proposed to calculate the phase shift caused by a non-trivial potential whose scaling is more than quadratic in position~\cite{ufrecht2020,Overstreet2021}.
Here, we use the perturbative methods developed in Ref.~\cite{ufrecht2020} and summarized in Appendix~\ref{sec_App_A}.
For species $i$, the total accumulated phase $\Phi_i$ can be decomposed as 
\begin{equation}
\label{eq_Phi_gen_giese}
    \Phi_i=\Phi_{i,\rm 0}+\Phi_{i,\rm pert}
\end{equation}
where $\Phi_{i,\rm 0}$ and $\Phi_{i,\rm pert}$ denote respectively the \textit{unperturbed phase} induced by a quadratic potential and the momentum transfer, as well as a \textit{perturbative} phase contribution.
In the following, we consider a polynomial potential of order $N$ of the form
\begin{equation}
\label{eq_V_poly}
    V_i(x)=\sum_{n=1}^N c_{i,n} x^n,
\end{equation}
which can be seen as an expansion of an arbitrary potential around the initial c.m. position of the atoms $x_0=0$.
In this study we only consider the contributions up to order $N=4$.
The coefficients $c_{i,n}$ also include an index for the species that account for species-dependent potentials discussed below.
Appendix~\Ref{sec_App_A} features the derivation of $\Phi_{i,\rm pert}$ for a perturbative potential for a single atomic species.
The contribution of terms beyond $N=4$ in lowest order can easily be obtained using Eq.~\eqref{eq:simplified-phase-giese-appendix}.
When working with different expansion coefficients, one has to check of course the required order of the perturbative expansion, as shown in the example given in Ref.~\cite{ufrecht2020}.

Since the species have different masses $m_i$, different effective wave numbers $k_i$, different interrogation times $T_i$, different initial positions $x_{0,i}$, velocities $v_{0,i}$, position widths $\sigma_{x,i}$ and velocity widths $\sigma_{v,i}$, we equip all quantities with index $i$ and find the phase
\begin{align}
\begin{split}
    \Phi_i = &- \frac{2k_i T_i^2}{m_i}c_{i,1} - 2\frac{2k_i T_i^2}{m_i} x_i(T_i) c_{i,2} + \kappa_i c_{i,3}\\
    &+ 4\left[\kappa_i x_i(T_i) + \dfrac{k_i T_i^2}{m_i}(4x^3_i(T_i)-2T_i^3 v_{0,i} \sigma_{v,i}^2)\right]c_{i,4},
\end{split}
\end{align}
with abbreviation 
\begin{equation}
    \kappa_i=- \dfrac{k_i T_i^2}{m_i}\left[6x_i^2(T_i)+6\sigma^2_{x,i}+T_i^2\left(v^2_{0,i}+\left(\dfrac{\hbar k_i}{m_i}\right)^2+ 7\sigma^2_{v,i}\right)\right],
\end{equation}
and $x_i(T_i)= x_{0,i} + v_{0,i} T_i$.

In the following we calculate the constraints on the expansion coefficients including their uncertainties by imposing  that the uncertainty in the differential acceleration for each order $n$ be below the target uncertainty i.e.
\begin{equation}
    \Delta \delta a^{(n)}  = \Delta\left( \dfrac{\Phi_A^{(n)}}{2k_A T_A^2}-\dfrac{\Phi_B^{(n)}}{2k_B T_B^2}\right) \leq \eta g_0 \,,
\end{equation}
for a self-gravity potential, black body radiation, as well as the second-order Zeeman effect. Note that we do not assign an uncertainty to $k_i$, $T_i$ and $m_i$ as they are known sufficiently well to not be limiting (see e.g. Sec.~\ref{sec:Acc_application}).

\subsubsection{Self gravity potential}\label{sec:systematic-effects-self-gravity}
An inhomogeneous distribution of the satellite's mass yields a gravitational potential inducing a spurious bias phase shift limiting the sensitivity to a possible WEP violation.
To study this effect we expand the gravitational potential of the satellite around a point corresponding to the nominal position $x_0=0$ of the c.m. of the BECs. 
For simplicity we only focus our analysis along the sensitive axis of the experiment, where the effect is largest. We write the Newtonian gravitational potential of a spherically symmetric source mass acting on the atoms as

\begin{equation}
\label{eq_V_SG}
    V_{i,\rm SG} = -\dfrac{G M m_i}{x_{M}}\left[1+\sum_{n=1}^{N}\left(\dfrac{x}{x_{M}}\right)^n\right],
\end{equation}
where $G$ is the gravitational constant, $M$ is the source-mass, $m_i$ is the mass of the atoms, $x$ is the position of the atoms and $x_{M}$ is the position of the source mass. 
Comparing Eq.~\eqref{eq_V_SG} to Eq.~\eqref{eq_V_poly}, one has, $c_{i,n}=-G M m_i/x^{n+1}_{M}$, $\forall n\in\mathbb{N}$. 

Tab.~\ref{tab_SG} provides constraints on the different self-gravity contributions and their uncertainties {\it that are synchronous with the signal}. 
Note that, as discussed in section \ref{sec:Adv_Sat}, the corresponding static constraints, i.e., the actual knowledge of the satellite's mass distribution, are up to 8 orders of magnitude less stringent.

\begin{table}[h!]
\caption{Maximum allowed self gravity variations and their maximum allowed uncertainties that are synchronous with the WEP violation signal, given in terms of $GM/x^{n+1}_M$ for $\eta=10^{-15}$ and $\eta=10^{-17}$. Note that the knowledge of the static self-gravity coefficients may be up to 8 orders of magnitude less stringent (cf. (i) and (ii) of Sec.~\ref{sec:Adv_Sat}).}
\begin{tabular}{c||c|c|c}
\hline \\ [-1.5ex]
n & $GM/x^{n+1}_M$\,($\eta=10^{-15}$) &  $GM/x^{n+1}_M$\,($\eta=10^{-17}$) & unit\\
\hline 
\hline
2 & $7.5 \times 10^{-10} \pm 2.1 \times 10^{-10}$ & $1.3 \times 10^{-11} \pm 3.4 \times 10^{-12}$  & s$^{-2}$\\
3 & $1.2\times 10^{12} \pm 6.4\times 10^{-13}$ & $2.3\times 10^{10} \pm 2.1\times 10^{-16}$ & m$^{-1}$.s$^{-2}$\\
4 & $9.5 \times 10^{-8} \pm 2.7 \times 10^{-8}$ & $5.1 \times 10^{-11} \pm 1.4 \times 10^{-11}$ & m$^{-2}$.s$^{-2}$\\
\hline
\end{tabular}
\label{tab_SG}
\end{table}

It should be emphasized here that the $n=2$ coefficient is a local gravity gradient which can be compensated the same way as the Earth's gravity gradient following the gravity gradient cancellation method discussed in Refs.~\cite{Roura2017a,loriani2020}.
 
\subsubsection{Black body radiation}
The effect of thermal radiation leads to black body radiation (BBR) acting as an extra external potential of the form~\cite{Haslinger2018},
\begin{equation}
\label{eq_V_BBR}
    V_{i, \rm BBR}(x)=\dfrac{2\alpha_i \sigma }{c \epsilon _0} T_{\rm tube}^4(x),
\end{equation}
where $\alpha_i$ is the static polarizability of atomic species $i$, $\sigma$ the Stephan-Boltzmann constant, $\epsilon_0$ the vacuum permittivity and $T_{\rm tube}(x)$ the temperature profile inside the vacuum tube at position $x$ along the sensitive axis.
To calculate the effect we expand $T_{\rm tube}(x)$ around $x_0=0$ analogously to Sec.~\ref{sec:systematic-effects-self-gravity}.
We write $T_{\rm tube}(x)=\sum_0^\infty t_n x^n$. 
Comparing Eq.~\eqref{eq_V_BBR} to Eq.~\eqref{eq_V_poly}, one has to leading order in $t_0$ the coefficients $c_{i,n}= 8\sigma \alpha_i t_0^3 t_n/(c \epsilon_0)$, $\forall n\in\mathbb{N}$.
The constraints on the temperature gradients {\it that vary synchronously with the signal} are given in Tab.~\ref{tab_BBR}.

\begin{table}[h!]
\caption{Requirements on the temperature gradients and their uncertainties that vary synchronously with the signal. Here we assume an average temperature of $t_0=283$\,K with uncertainty $\Delta t_0=1$\,mK and an average temperature gradient of $t_1 = 5$\,mK/m~\cite{Touboul2017}. Numerical values have been obtained for a  static polarizability of the atoms: $\alpha_{Rb}=2\pi\hbar\times 0.0794 \times 10^{-4} $\,Hz.V$^{-1}$m$^2$ and $\alpha_K=\alpha_{Rb}/1.1$~\cite{Vanier1989}. Note that the constraints on the purely orbital component may be up to 3 orders of magnitude less stringent (cf. (ii) of Sec.~\ref{sec:Adv_Sat}).}
\begin{tabular}{c||c|c|c}
\hline \\ [-1.5ex]
n & $t_n$\,($\eta=10^{-15}$) & $t_n$\,($\eta=10^{-17}$) & unit	\\
\hline 
\hline
1 & $\pm 2.5\times 10^{-5}$ & $\pm 2.5\times 10^{-7}$ & K.m$^{-1}$ \\
2 & $3.6 \pm 1.0$ & $6.1\times 10^{-2} \pm 1.6\times 10^{-2}$ & K.m$^{-2}$ \\
3 & $1.3 \times 10^6 \pm 2.1\times 10^{-3}$ & $3.4\times 10^{4} \pm 6.8\times 10^{-7}$ & K.m$^{-3}$ \\
4 & $3.1\times 10^2 \pm 1.1\times 10^2$ & $1.6 \pm 5.6\times 10^{-2}$ & K.m$^{-4}$ \\ 
\hline
\end{tabular}
\label{tab_BBR}
\end{table}

Note that the onboard temperature gradients are expected to vary mainly at orbital frequency and its harmonics. Because of the phase modulation of the signal by controlled rotations, as discussed in point (ii) in Sec.~\ref{sec:Adv_Sat}, the error in the knowledge of the amplitude of that variation could be up to three orders of magnitude less stringent than the constraints given in Tab.~\ref{tab_BBR}.

\subsubsection{Second-order Zeeman effect}

We consider an interferometer sequence operated with atoms in the $m_F=0$ state and is thus up to first order insensitive to magnetic effects.
Here, we study the impact of the second-order Zeeman effect on the differential phase shift~\cite{Vanier1989}.
The potential induced by the presence of magnetic field can be written as 
\begin{equation}
V_{i,\rm B}(z)= \pi \hbar \chi_i B_{\rm tube}^2(x),
\end{equation}
where $\chi_i$ is the second-order Zeeman coefficient of atomic species $i$ and $B_{\rm tube}(x)$ is the magnetic field inside the vacuum tube at position $x$ along the sensitive axis.
We evaluate the effect of the magnetic field gradients the same way as in the previous sections, i.e. we expand the magnetic field in a series expansion $B_{\rm tube}(x)=\sum_0^\infty b_{n}x^n$ and calculate constraints on the coefficients $b_{n}$. 
The constraints on the magnetic field gradients {\it that vary synchronously with the signal} are given in Tab.~\ref{tab_Z}.

\begin{table}[h!]
\caption{Requirements on the magnetic field gradients and their uncertainties that vary synchronously with the signal. We assume here $b_0=100$\,nT with uncertainty $\Delta b_0 = 50$\,pT and $b_1 = 6$\,nT/m. From Ref.~\cite{Vanier1989} we have $\chi_{Rb} = 575.14\times 10^8\, \rm Hz/T^2$ and $\chi_{K} = 15460\times 10^8\, \rm Hz/T^2$. Note that the knowledge of the main component at 2$f_{orb}$  may be up to 3 orders of magnitude less stringent (cf. (ii) of Sec.~\ref{sec:Adv_Sat}) and can be measured independently of the signal (cf. (iii) of Sec.~\ref{sec:Adv_Sat}).}
\begin{tabular}{c|c|c|c}
\hline \\ [-1.5ex]
n & $b_n$\,($\eta=10^{-15}$) & $b_n$\,($\eta=10^{-17}$) & unit	\\
\hline 
\hline
1 & $\pm 2.2\times 10^{-3}$ & $\pm 2.2\times 10^{-5}$ & nT.m$^{-1}$ \\
2 & $9.8\times 10^2 \pm 2.8\times 10^{2}$ & $1.6 \times 10^1 \pm 4.5$ & nT.m$^{-2}$ \\
3 & $1.1\times 10^8 \pm 3.4\times 10^{-1}$ & $3.0\times 10^6 \pm 1.1\times 10^{-4}$ & nT.m$^{-3}$ \\
4 & $6.6\times 10^6 \pm 2.2\times 10^{4}$ & $1.8\times 10^{5} \pm 1.1\times 10^{1}$ & nT.m$^{-4}$ \\ 
\hline
\end{tabular}
\label{tab_Z}
\end{table}

In the case of a circular orbit, the main time variation of $B_{\rm tube}^2(x)$ will be at $2f_{\rm orb}$ because of the dipolar nature of the Earth's magnetic field, and thus decorrelate well from the EP-violating signal at $f_{\rm orb}$. 
The effect can therefore be modelled and subtracted (cf. (ii) of Sec.~\ref{sec:Adv_Sat}) additionally to the reduction by about 3 orders of magnitude because of the phase modulation of the signal by controlled rotations (point (ii) in Sec.~\ref{sec:Adv_Sat}).

We emphasize here that magnetic field gradients below the nT/m level \cite{Wolf2006a} are achieved on $30$~cm scales on the ground, in a much more perturbed magnetic environment than in space, and at a few nT/m over 10~m scales~\cite{Asenbaum2020,Wodey2020}.

\section{Constraints on the spacecraft}\label{sec:spacecraft-constraints}
Accelerations and rotations of the satellite can directly translate into additional phase shifts in the interferometric measurement. We now derive the corresponding requirements on spurious accelerations and rotations of the spacecraft.

\subsection{Acceleration}\label{sec:acceleration}
\subsubsection{Systematic effect}
Any common accelerations of the two test masses can be suppressed by carefully choosing the experimental parameters as well as combining the phase shifts for each species in an optimal way (see Eq.~\eqref{phi_gen}).
An obvious choice would be $\mathcal{A}=1$ and $\mathcal{B}=-1$ leading to the direct subtraction of both phases. 
However, a more rigorous analysis reveals that we can exploit this freedom to construct a differential phase shift observable that is non-sensible to common accelerations. 
The transfer function of the differential atom interferometer phase, $\Phi_{\rm gen}$ in Eq.~\eqref{phi_gen}, defining the response of the interferometer with respect to vibrational noise is given by \cite{cheinet2008}
\begin{equation}
    H(\omega) = - 4i \left[2\mathcal{A}k_A \sin ^2 (\omega T_A/2) +  2\mathcal{B}k_B\sin^2 (\omega T_B/2)\right].
\end{equation}

The response to common accelerations can be reduced to zero by setting $T = T_A = T_B$ and $\mathcal{A} k_A = - \mathcal{B} k_B$. To keep $\mathcal{A}\approx-\mathcal{B}\approx 1$, we choose $\mathcal{A} = 2k_B/(k_A+k_B)$ and $\mathcal{B} = -2k_A/(k_A+k_B)$ such that
\begin{equation}
    \label{eq_AI_phase2}
    \Phi_{\rm gen} = \frac{2k_B}{k_A+k_B}\Phi_A - \frac{2k_A}{k_A+k_B}\Phi_B.
\end{equation}

Rewriting the accelerations as $a_i=a_{\rm c}\pm a_{\rm nc}\pm \eta g_0/2$ where $a_{\rm c}$ ($a_{\rm nc}$) encompasses all the common (non-common) accelerations between the two species and where $a_\eta$ is the extra acceleration due to a violation of the WEP, one has:
\begin{equation}
    \Phi_{\rm gen} = \frac{4k_A k_B}{k_A+k_B} (2 a_{\rm nc} + \eta g_0) T^2.
\end{equation}
The combination is insensitive to any common accelerations $a_{\rm c}$, but the sensitivity to $\eta g_0$ stays approximately untouched assuming $k_A\approx k_B$.

Of course, the performance of the \textit{acceleration free combination} defined by Eq.~\eqref{eq_AI_phase2} relies on the exact knowledge of the wave numbers $k_i$. 
Assuming that this knowledge is limited up to an uncertainty $\Delta k_i$, leads to an uncertainty in the differential phase of
\begin{equation}
\begin{split}
    \Delta\delta\Phi_{\rm gen} = 4\frac{k_B\Delta k_A - k_A\Delta k_B}{k_A+k_B} a(t) T^2,
\end{split}
\end{equation}

where $a(t) = a_A(t) = a_B(t)$ is the residual acceleration of the satellite common to both species. We require that the component of $\Delta\delta\Phi$ that is modulated with orbital frequency stays below the Eötvös signal. Thus, we find
\begin{equation}
     |\Delta k_A/k_A - \Delta k_B/k_B| \times  a(t)\vert_{\omega_{\rm orb}} \leq \eta g_0.
     \label{eq_wavenumber_req}
\end{equation}
leaving a requirement on the relative knowledge $\Delta k_i/k_i$ for a given $a(t)\vert_{\omega_{\rm orb}}$. Note that the laser frequency is well known and the uncertainty $\Delta k_i$ is dominated by pointing errors \cite{schubert2019scalable}.

\subsubsection{Acceleration noise}
\begin{figure}[h!]
\includegraphics[width=\linewidth]{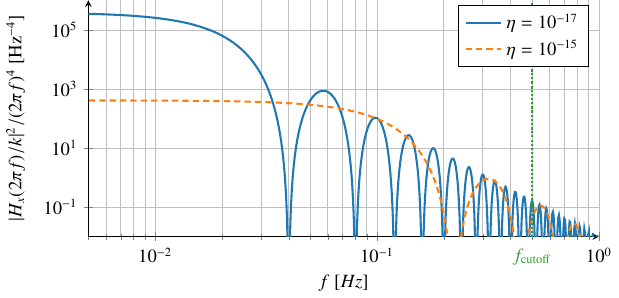}
\caption{Linear acceleration transfer function, $H_x(\omega)/\omega^2 = 8 k \sin^2(\omega T/2) / \omega^2$, Eq.~\eqref{eq:acceleration-noise-req}, for the ${}^{87}$Rb atom interferometer using the parameters in Tab.~\ref{tab:params}. It shows constant behavior for low frequencies and drops steeply $\sim f^{-2}$ at high frequencies $f \gg 2/T$. The frequency cutoff $f_{\rm cutoff}$ up to which the integration in Eq.~\eqref{eq:acceleration-noise-req} is performed is marked in dotted green.}
\label{fig:linear-acc-transfer-func}
\end{figure}

Although any common accelerations are suppressed by the \textit{acceleration free combination}, Eq.~\eqref{eq_AI_phase2}, we require that the phase shift induced by acceleration noise stays within a certain region around mid-fringe, i.e., the point of maximum phase sensitivity. 
We quantify this requirement by setting $\Delta\Phi_i^a \leq \pi/10$ where $\Delta\Phi_i^a$ is the uncertainty in the phase shift due to acceleration noise with power spectral density $S_a(\omega)$ \cite{cheinet2008},
\begin{equation}
    (\Delta\Phi_i^a)^2 = \int_0^{+\infty} \frac{S_a(\omega)}{\omega^4}\left|8 k_i \sin ^2 \left(\dfrac{\omega T}{2}\right)\right|^2 d\omega \leq \left( \frac{\pi}{10}\right)^2.
    \label{eq:acceleration-noise-req}
\end{equation}
The integration can be restricted to an area $ \omega \in [2\pi/T_c, 2\pi f_\mathrm{cutoff}]$ where $T_c$ is the cycle time of the measurement. 
At high frequencies the linear acceleration transfer function drops steeply as shown in Fig.~\ref{fig:linear-acc-transfer-func}, so we can limit the integration up to a certain cutoff $f_\mathrm{cutoff}$. 
At low frequencies, $\Delta\Phi_a$ reduces to the variance of the atom-interferometer phase $2k_i \langle a\rangle T^2$ where $\langle a \rangle$ is the average acceleration during the interrogation time of the atoms.
We emphasize here that the slowly varying acceleration noise can be approximated by a low-order polynomial and used to suppress the noise in following measurements by feed forwarding the information to the laser frequency. 
In order to stay at mid-fringe, the change in acceleration in-between cycles should follow the inequality,
\begin{equation}
    \langle\dot\Phi_i\rangle = 2k_iT^2  \langle\dot a \rangle T_c \leq \pi/10 ,
    \label{eq:mean-linear-acc-req}
\end{equation}
where $\langle\dot\Phi_i\rangle$ is the average change of the interferometer phase from one cycle to the next.

\subsubsection{Application} \label{sec:Acc_application}

The \textit{acceleration free combination}, Eq.~\eqref{eq_AI_phase2}, ensures that any accelerations that are common to both species are suppressed, leaving only the requirement on the knowledge of the relative wave numbers, Eq.~\eqref{eq_wavenumber_req}. Taking the drag-free controlled MICROSCOPE satellite as an example \cite{robert2022microscope}, no residual acceleration exceeding $a(t)\vert_{\omega_{\rm orb}} > 10^{-12}$ m/s$^2$ was observed, implying that $(\Delta k_A/k_A - \Delta k_B/k_B) \leq 6.6 \times 10^{-5}$ is sufficient to reach $\eta \leq 10^{-17}$, well within the reach of present day laser systems.

The mid-fringe requirement leads to requirements on the single-species atom interferometer acceleration noise. Tab.~\ref{tab:acc-req} features the constraints evaluated for the parameters in Tab.~\ref{tab:params} assuming white acceleration noise and using $f_\mathrm{cutoff} = 0.5$~Hz.

\begin{table}[h!]
\caption{Constraints on spurious linear accelerations of the spacecraft for $\eta=10^{-15}$ and $\eta=10^{-17}$.}
\begin{tabular}{c|c|c|c|c}
\hline \\ [-1.5ex]
Quantity & eq. & $\eta=10^{-15}$ & $\eta=10^{-17}$ & Unit	\\
\hline 
\hline
$\sqrt{S_a}$ & \eqref{eq:acceleration-noise-req} & $3.5\times 10^{-9}$ & $4.1\times 10^{-10}$ & m/s$^{2}$/$\sqrt{\mathrm{Hz}}$ \\
$\langle\dot a\rangle$ & \eqref{eq:mean-linear-acc-req} & $3.2\times 10^{-11}$ & $2.6\times 10^{-13}$ & m/s$^{3}$ \\
\hline
\end{tabular}
\label{tab:acc-req}
\end{table}

\subsection{Rotation}
Systematic and statistical effects due to rotations are not suppressed by the \textit{acceleration free combination}, Eq.~\eqref{eq_AI_phase2}, and, thus, are of particular interest when setting constraints on the platform.

\begin{figure}[h!]
\includegraphics[width=\linewidth]{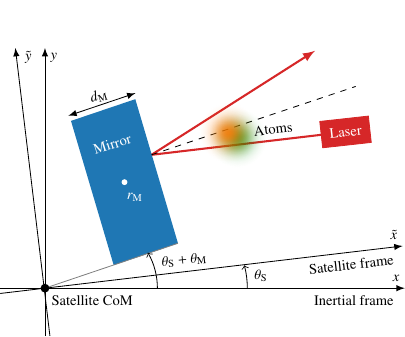}
\caption{Schematic representation of the experimental setup inside the satellite. Initially, the inertial frame coincides with the satellite frame. After release, the satellite undergoes rotations $\theta_\text{S}(t)$ whereas the mirror (blue) is rotated by $\theta_\text{M}(t)$. The laser head's position (red) is fixed in the satellite frame. The $x$-axis of the inertial frame is chosen to be initially parallel to the sensitive axis of the interferometer (dashed). The incident and reflected laser beam is marked as a red line. The atoms' position is marked in green and orange.}
\label{fig:rotation-setup}
\end{figure}

We now investigate the configuration shown in Fig.~\ref{fig:rotation-setup} to derive the phase shifts induced by rotations of the mirror or the spacecraft. We work in an inertial frame whose origin coincides with the center of mass of the satellite for all times $t$. The orientation is chosen such that the sensitive axis of the interferometer is aligned with the $x$-axis when the atoms are released.
The mirror is assumed to be rectangular with a thickness of $d_{\rm M}$ with its center of mass positioned at $\bf{r_{\rm M}}$.
Small rotations of the mirror $\theta_{\rm M}$ simply add to the effect of the rotation of the satellite $\theta_{\rm S}$ such that the overall rotation can be defined as $\theta(t) = \theta_{\rm M}(t) + \theta_{\rm S}(t)$.

\subsubsection{Systematic effect}

Spurious rotations of the spacecraft cause additional accelerations scaling with the initial kinematics of the atoms.
We first derive constraints on the angular velocity of the satellite by looking at the case of a constant rotation rate $\Omega$, $\theta(t) = \Omega t$ around the $z$-axis (see Fig.~\ref{fig:rotation-setup}). 

Using geometric considerations, the atom interferometer phase can be derived (see Appendix.~\ref{sec:rotation-derivation}):
\begin{equation}
    \Phi_i = 4 k_i v_{y,0,i} \Omega T^2 + 2 k_i(r_{x,0,i}  - v_{x,0,i} (T + T_0)) (\Omega T)^2,
\label{eq:angular-velocity-ati-phase}
\end{equation}
to second order in $\Omega T$. 
Here $T_0$ is the dead time, corresponding to the time between release and first laser pulse, and $r_{a,0,i}$ ($v_{a,0,i}$) denotes the atoms' initial position (velocity) in $a\in \{x,y,z\}$ direction.

When calculating the differential phase according to Eq.~\eqref{eq_AI_phase2}, the dependencies on the individual initial kinematics directly translate into dependencies on the differential initial kinematics. 
Any systematical uncertainty must be below the target signal at $\omega_{\rm orb}$. 
Thus, we find the following requirement on the satellite's angular velocity at orbital frequency
\begin{equation}
    \Omega\rvert_{\omega_{\rm orb}} \leq \eta g / (2 \delta v_{y,0}),
    \label{eq:angular-velocity-sys-req}
\end{equation}
for the first order in Eq.~\eqref{eq:angular-velocity-ati-phase}. The second order leads to a requirement on $\Omega^2$ at orbital frequency
\begin{equation}
\begin{split}
    \Omega^2\rvert_{\omega_{\rm orb}} &\leq \eta g / \delta r_{x,0}, \\
    \Omega^2\rvert_{\omega_{\rm orb}} &\leq \eta g /(\delta v_{x,0} (T+T_0)),
\end{split}
\label{eq:angular-velocity-squared-sys-req}
\end{equation}
which translates into a requirement on $\Omega$ at half the orbital frequency: $\Omega\rvert_{\omega_{\rm orb}/2} = \sqrt{\Omega^2\rvert_{\omega_{\rm orb}}}$ and other cross terms of the form $\Omega\rvert_{\omega_1}\Omega\rvert_{\omega_2}$, where $|\omega_1 \pm \omega_2| = \omega_{\rm orb}$ \footnote{Suppose the angular velocity is modulated according to $\Omega\rvert_{\omega_1} = \Omega_1 \cos (\omega_1 t)$ and $\Omega\rvert_{\omega_2} = \Omega_2 \cos (\omega_2 t)$ where $|\omega_1\pm\omega_2|=\omega_{\rm orb}$. Then it follows that $\Omega\rvert_{\omega_1}\Omega\rvert_{\omega_2} = \Omega_1\Omega_2 [\cos ((\omega_1-\omega_2)t) + \cos ((\omega_1+\omega_2)t)]/2$. In particular, $|\omega_1 \pm \omega_2| = \omega_{\rm orb}$ leads to $\Omega\rvert_{\omega_1}\Omega\rvert_{\omega_2} = (\Omega^2)\rvert_{\omega_{\rm orb}}$.}.

\subsubsection{Rotation noise}
In the following, we consider arbitrary satellite rotations $\theta(t)$ to give constraints on rotational noise and derive further suppression techniques. 

Up to first order in $\theta(t)$, the atom interferometer phase is given by (see Appendix~\ref{sec:rotation-derivation}):
\begin{equation}
\begin{split}
    \Phi_i = 2 k_i [&
        \tilde{r}_{y,i}(T_0) \theta (-T) - 2 \tilde{r}_{y,i}(T_0+T)\theta(0) \\ 
        & + \tilde{r}_{y,i}(T_0 + 2T)\theta(T)],
\end{split}
\label{eq:rotations-single-ati-phase-first-order}
\end{equation}
where $\tilde{r}_{y,i}(t) = r_{y,0,i} + (v_{y,0,i} + r_{x,0,i} \Omega_0) t$ denotes the atom's position in the rotated reference frame with $\Omega_0 = d\theta/dt\vert_{t=0}$ and $t$ is the time after release.

The corresponding transfer function is obtained by taking the Fourier transform of the sensitivity function associated to the signal of Eq.~\eqref{eq:rotations-single-ati-phase-first-order},
\begin{equation}
\begin{split}
    H_\theta (\omega) 
    &= 4 k_{AB} [ -2 i (\delta r_{y,0} + \delta v_{y,0} (T_0 + T)) \sin^2 (\omega T /2) \\
    &\phantom{= 4 k_{AB} [} + \delta v_{y,0} T \sin (\omega T) ],
    \label{eq:rot-transfer-func}
\end{split}
\end{equation}
where $k_{AB} = 2k_Ak_B/(k_A+k_B)$.
\begin{figure}[tb]
    \centering
    \includegraphics[width=\linewidth]{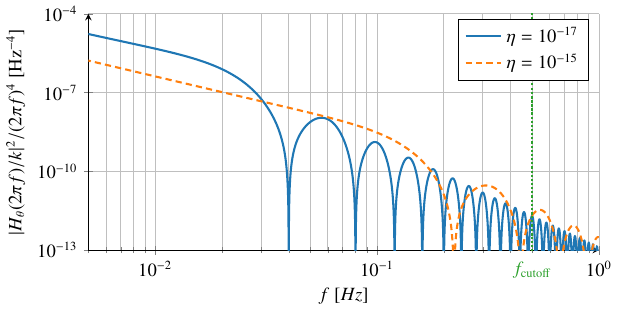}
    \caption{Angular acceleration transfer function, Eq.~\eqref{eq:rot-transfer-func}, using the parameters in Tab.~\ref{tab:params}. For low frequencies $f \ll 2/T$ it behaves as $f^{-1}$ while for larger frequencies $f \gg 2/T$ it follows $f^{-2}$, similar to the linear acceleration transfer function (see Fig.~\ref{fig:linear-acc-transfer-func}). The frequency cutoff $f_{\rm cutoff}$ up to which the integration in Eq.~\eqref{eq:rotation-noise-unc} is performed is marked in dotted green.}
    \label{fig:angular-acceleration-transfer-func}
\end{figure}
The uncertainty of the differential phase induced by rotations $\theta$ of the satellite is then given by
\begin{equation}
    (\Delta \delta\Phi_{\ddot{\theta}})^2 = \int_0^{+\infty} |H_\theta (\omega)|^2 \frac{S_{\ddot{\theta}}}{\omega^4} (\omega) d\omega,
    \label{eq:rotation-noise-unc}
\end{equation}
where $S_{\ddot{\theta}}(\omega)$ is the power spectral density of angular acceleration noise.

We require that any uncertainty in the differential phase induced by rotational noise of the satellite per cycle must be below the shot noise limit (see Sec.~\ref{sec:statistical-error}):
\begin{equation}
    \Delta \delta\Phi_{\ddot{\theta}} \leq \Delta \delta\Phi_\mathrm{SN} = \frac{2\sqrt{k_B^2 + k_A^2}}{k_A+k_B}\frac{1}{\sqrt{N}} \approx \sqrt{\frac{2}{N}}.
    \label{eq:rotation-noise-req}
\end{equation}
We can restrict the integration in Eq.~\eqref{eq:rotation-noise-unc} to $\omega \in [2\pi/T_c, 2\pi f_\mathrm{cutoff}]$ as for high frequencies the transfer function drops steeply (see Fig.~\ref{fig:angular-acceleration-transfer-func}). For low frequencies $\omega < 2\pi/T_c$ the uncertainty reduces to the Coriolis phase $4 k_{AB} \delta v_{y,0} \langle \Omega \rangle T^2$ where $\langle\Omega\rangle$ is the average value of $\dot\theta$ during the atom interferometer sequence. This can be treated as a systematic effect resulting in a requirement on $\langle\Omega\rangle$ (see Eq.~\eqref{eq:angular-velocity-sys-req}).

Additional noise constraints arise from the \textit{mid-fringe requirement} and from the coupling of the atoms' finite velocity spread to $\langle\Omega\rangle$ inducing shot to shot noise. The requirement to stay at mid-fringe is treated the same way as low frequency acceleration noise by feed forwarding the results of previous measurements. This yields a requirement on the average phase change per cycle,
\begin{equation}
    \langle\dot\Phi\rangle = 4k_i T^2 v_{i,y,0}\dot{\langle\Omega\rangle} T_c \leq \pi/10.
    \label{eq:mean-angular-acc-req}
\end{equation}
Additional phase noise arises due to the limited knowledge of the atoms' kinematics. The shot to shot phase noise induced by the position and velocity uncertainty $\Delta r_{i,0}$ and $\Delta v_{i,0}$ is required to be smaller than the shot noise,
\begin{equation}
    \left( 4k_i T^2 \frac{\sigma_{v,i}}{\sqrt{N}}\langle\Omega\rangle \right)^2 + \left( 2 k_i T^2 \frac{\sigma_{r,i}}{\sqrt{N}}\langle\Omega^2\rangle \right)^2 \leq \frac{1}{N},
    \label{eq:mean-angular-vel-req}
\end{equation}
assuming a shot noise limited process of determining the atoms' mean position and velocity: $\Delta r_{i,0} = \sigma_{r,i}/\sqrt{N}$ and $\Delta v_{i,0} = \sigma_{v,i}/\sqrt{N}$ \cite{hensel2021}.

\subsubsection{Application}

Tab.~\ref{tab:rot-req} lists the constraints coming from rotations of the spacecraft including their evaluation using the parameters of Tab.~\ref{tab:params} assuming white angular acceleration noise and using $f_\mathrm{cutoff} = 0.5$~Hz in Eq.~\eqref{eq:rotation-noise-unc}. Note that, in particular for $\Omega\rvert_{\omega_{\rm orb}}$ and $\Omega\rvert_{\omega_{\rm orb}/2}$, these requirements do not take into account the phase modulation of the signal by controlled rotations (point (ii) of Sec.~\ref{sec:Adv_Sat}) and thus could be relaxed by about 3 orders of magnitude.

\begin{table}[h!]
\caption{Constraints on angular velocity for $\eta=10^{-15}$ and $\eta=10^{-17}$. Note that, in particular for $\Omega\rvert_{\omega_{\rm orb}}$ and $\Omega\rvert_{\omega_{\rm orb}/2}$, these requirement could be relaxed by about 3 orders of magnitude when applying phase modulation as described in point (ii) of Sec.~\ref{sec:Adv_Sat}.}
\begin{tabular}{c|c|c|c|c}
\hline \\ [-1.5ex]
Quantity & eq. & $\eta=10^{-15}$ & $\eta=10^{-17}$ & Unit	\\
\hline 
\hline
$\Omega\rvert_{\omega_{\rm orb}}$ & \eqref{eq:angular-velocity-sys-req} & $3.3\times 10^{-9}$ & $3.3\times 10^{-10}$ & rad/s \\
$\Omega\rvert_{\omega_{\rm orb}/2}$ & \eqref{eq:angular-velocity-squared-sys-req} & $3.8\times 10^{-5}$ & $5.1\times 10^{-6}$ & rad/s \\
$\sqrt{S_{\ddot{\theta}}}$ & \eqref{eq:rotation-noise-req} & $8.4\times 10^{-6}$ & $3.2\times 10^{-7}$ & rad/s$^{2}$/$\sqrt{\mathrm{Hz}}$ \\
$\dot{\langle\Omega\rangle}$ & \eqref{eq:mean-angular-acc-req} & $1.6\times 10^{-5}$ & $1.3\times 10^{-7}$ & rad/s$^{2}$ \\
$\langle\Omega\rangle$ & \eqref{eq:mean-angular-vel-req} & $7.5\times 10^{-6}$ & $5.4\times 10^{-7}$ & rad/s \\
$\langle\Omega^2\rangle$ & \eqref{eq:mean-angular-vel-req} & $1.5\times 10^{-5}$ & $9.8\times 10^{-8}$ & rad$^2$/s$^{2}$ \\
\hline
\end{tabular}
\label{tab:rot-req}
\end{table}

\subsection{Orbit control}\label{sec:orbit-control}

The error mitigation techniques as well as the extraction of the target signal rely on the knowledge of the local gravitational potential. Since the inertial quantum sensor performs a local differential measurement, orbit errors only play a role via perturbing effects from external factors. An error in the knowledge of the satellite's position at the time of a measurement directly translates to an error due to an incorrect estimation of the corresponding differential gravitational acceleration and its gradients. 

\subsubsection{Model}
To analyze the effect of an orbit uncertainty we utilize a satellite simulator which enables us to study the effects of statistical and systematic uncertainties in the satellite's orbit or attitude. The simulator, called SQUID (Satellite-based QUantum systems for Inertial sensing and Discovery of new physics), allows to synthetically generate a space-borne atom interferometer signal and also analyze it assuming arbitrary orbit and attitude configurations. Here, we generate a realistic signal using a distorted orbit and fit it using a model assuming a perfect, i.e., circular, orbit to study orbit uncertainty induced limitations. 

We implement orbit distortions using the Hill model which characterizes the position errors (for weakly eccentric orbits) at time $t$ according to \cite{duchayne2009}
\begin{equation}
\begin{split}
    \Delta R(t) &= \tfrac{1}{2} X \cos (\omega_{\rm orb} t + \varphi_R) + c_R , \\
    \Delta T(t) &= - X \sin (\omega_{\rm orb} t + \varphi_R) - \tfrac{3}{2}\omega_{\rm orb} c_R t + d_R , \\
    \Delta N(t) &= Y \cos (\omega_{\rm orb} t + \varphi_N) ,
\end{split}
\label{eq:hill-model}
\end{equation}
where ($\Delta R$, $\Delta T$, $\Delta N$) denotes the uncertainty in the (radial, tangential, normal) axis. $X$, $Y$, $c_R$ and $d_R$ are amplitude coefficients. For example, a radial uncertainty with $X>0$ relates to an eccentricity of $e = \sqrt{X/r}$ where $r$ is the radius of the inertial circular orbit.
A deviation from the circular orbit, e.g., one leading to $e>0$, introduces additional components of the gravity gradient that are modulated with the orbital frequency \cite{loriani2020}. 

The signal under consideration is given by
\begin{equation}
    \delta\Phi (t_j) = 2 [\eta g_x(t_j) + \delta r_{x,0} \Gamma_{xx}(t_j) ] k_{AB} T^2 ,
    \label{eq:ATI-signal-analysis}
\end{equation}
where $\Gamma_{xx}$ is the $xx$-component of the gravity gradient and $x$ equals the direction of the sensitive axis of the interferometer assuming an inertial attitude. The signal is sampled at certain satellite positions $r_S(t_j)$ where $t_j$ marks the times a measurement is performed: $t_j \in [0,\,T_c,\, \dots,\, T_{\rm sc}]$.

To perform the analysis, a signal is generated according to Eq.~\eqref{eq:ATI-signal-analysis} using positions $r_S$ computed for a circular orbit with a distortion given by Eq.~\eqref{eq:hill-model}. To estimate the uncertainty introduced by the distortion, we perform a least squares analysis of this signal using a fit model assuming an undistorted circular orbit (see Appendix~\ref{appendix:squid}).

\subsubsection{Application}
\begin{figure}[htb]
    \centering
    \includegraphics[width=\linewidth]{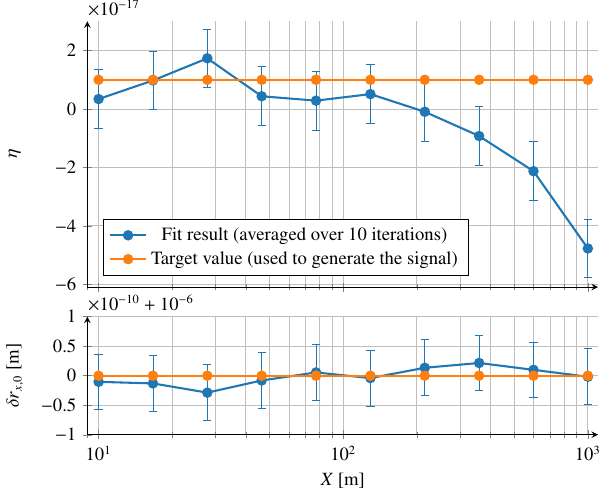}
    \caption{Orbit control analysis. The signal was generated using $\eta=10^{-17}$ and $\delta r_{x,0} = 1$ $\mu$m. The fit was performed for various distortion strengths $X$ (see Eq.~\eqref{eq:hill-model}) while every other parameter in Eq.~\eqref{eq:hill-model} was set to zero. Each blue point corresponds to 10 fits with different white noise that have been averaged. The error bars represent the standard deviation of the 10 fits.}
    \label{fig:orbit-control-analysis}
\end{figure}

To estimate the maximum allowed orbit distortion for the STE-QUEST mission proposal, we initialize the unperturbed orbit as circular with the parameters stated in Tab.~\ref{tab:params} ($\eta = 10^{-17}$). We will focus on $X$ because distortions along the radial and tangential axis directly couple into the signal through the gravity gradient. The normal axis is always perpendicular to the sensitive axis and will lead to less stringent requirements. The simulation was carried out with input values of $\eta=10^{-17}$ and $\delta r_{x,0} = 1$ $\mu$m. The result of the analysis, i.e., the resulting fitted values and uncertainties of $\eta$ and $\delta r_{x,0}$ as a function of $X$, is depicted in Fig.~\ref{fig:orbit-control-analysis}. 
It is clearly visible that the correct value for $\delta r_{x,0}$ is recovered in the fit with an uncertainty of $\pm 0.05$ nm, independent of the orbit error $X \leq 10^3$ m. The resulting $\eta$, however, drifts away from the expected value $\eta = 10^{-17}$ for increasing orbit errors $X$. After $X \approx 250\,\rm m$, the expectation value leaves the confidence interval of the fit. This would correspond to a requirement on the maximum tolerable eccentricity, $e \approx 5.6 \times 10^{-3}$. 

Note that this analysis does not take any attenuation techniques into account. Normally, the satellite's position is measured together with the differential acceleration leading to a requirement only on the knowledge of the orbit's eccentricity.

\subsection{Attitude control} 
\label{sec_attitude_control}
The de-correlation technique (see Sec.~\ref{sec:Adv_Sat}) to decouple the signal at interest, i.e., $\eta g_x(t)$, from spurious effects modulated at the same frequency relies on the control of the satellite's attitude. 
By periodically performing discrete rotations of the satellite during the science time, we introduce phase jumps in the Eötvös signal $\eta g_x(t)$ that help to de-correlate it from external influences modulated at orbital frequency that are not affected by these rotations. These rotations, however, need to be controlled up to a certain level to not introduce additional systematics reducing the sensitivity of the sensor. Here, we exploit SQUID to analyze uncertainties in these satellite rotations and set requirements on the attitude control system.

\subsubsection{Model}

For the numerical analysis, we proceed similarly as in Sec.~\ref{sec:orbit-control}. Here, the synthetic signal includes the WEP violation plus some spurious accelerations,
\begin{equation}
    \delta \Phi(t_j) = 2\left[\eta g_x(t_j) + \delta a_{\rm DC} + \delta a_{\rm orb}(t_j)\right] k_{AB}T^2,
    \label{eq:signal-de-correlation-analysis}
\end{equation}
where $t_j \in [0,\,T_c,\, \dots,\, T_{\rm sc}]$ denotes the times a measurement is performed. The linear gravitational acceleration $g_x(t_j) = g_x[r_S(t_j), \theta(t_j)]$ is determined by the satellite's position $r_S$ and attitude $\theta$ at time $t_j$. $\delta a_{\rm DC}$ denotes a spurious constant differential acceleration while $\delta a_{\rm orb}(t_j)$ is modulated at orbital frequency, i.e., $\delta a_{\rm orb}(t_j) = \delta a_{\rm orb, max} \cos (\omega_{\rm orb} t_j)$. Note that both of these additional differential accelerations are assumed to be immune to changes in the satellite's attitude.

Here, we focus on a circular orbit where the satellite is kept inertial but rotated by $10^\circ + \Delta\theta_{m}$ every 50 orbits. 
$\Delta\theta_m$ denotes the rotation noise that is drawn from a Gaussian distribution with zero mean and standard deviation $\sigma_{\Delta\theta}$ every time the satellite is rotated. To this signal, we additionally add atomic shot noise. 
The model matrix is constructed as in Eq.~\eqref{eq:ATI-model-analysis} assuming rotations of the satellite without any noise: $\Delta\theta_m=0$, $\forall m$.
Finally, we fit the signal using our model for the free parameters $\eta$, $\delta a_{\rm DC}$ and $\delta a_{\rm orb, max}$ for various noise levels $\sigma_{\Delta\theta}$.

\subsubsection{Application}
\begin{figure}[htb]
    \centering
    \includegraphics[width=\linewidth]{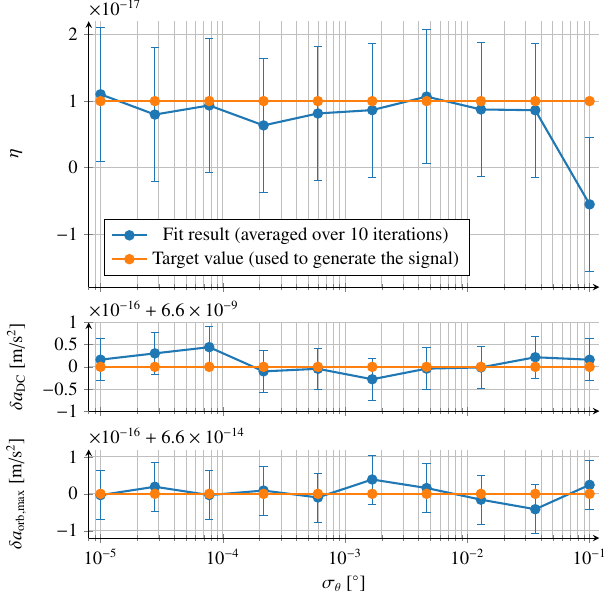}
    \caption{Attitude control analysis. The signal was generated using Eq.~\eqref{eq:signal-de-correlation-analysis} for $\eta = 10^{-17}$, $\delta a_{\rm DC} = 6.6 \times 10^{-9}$ m/s$^2$ and $\delta a_{\rm orb, max} = 6.6 \times 10^{-14}$ m/s$^2$. Each row displays the fit result for the respective parameter that is shown for various attitude noise levels $\sigma_{\Delta\theta}$. The orange points show the target values, whereas the blue points show the recovered value from the fit. The error bars represent the standard deviation of the 10 fits.}
    \label{fig:attitude-control-analysis}
\end{figure}

In the following, we will consider the parameters of STE-QUEST (see Tab.~\ref{tab:params}, $\eta=10^{-17}$) to set a requirement on the attitude control system with a focus on the de-correlation technique defined in Sec.~\ref{sec:Adv_Sat}.
The result of this analysis is depicted in Fig.~\ref{fig:attitude-control-analysis}. 
Here, we iterate over different rotation noise strengths $\sigma_{\Delta\theta}$ and try to recover the Eötvös parameter, $\eta$, a constant differential acceleration, $\delta a_{\rm DC}$ and a differential acceleration modulated at orbital frequency, $\delta a_{\rm orb,max}$, from the noisy signal. 
The signal is generated using $\eta = 10^{-17}$, $\delta a_{\rm DC} = g_0 \times 10^{-9}$ and $\delta a_{\rm orb, max} = g_0 \times 10^{-14}$ as these correspond to the orders of magnitude of uncertainties in the differential acceleration induced by black body radiation and magnetic fields (DC effects) and by self-gravity gradients (AC effect) (see Sec.~\ref{sec:systematic-effects}). 
The differential acceleration signals $\delta a_{\rm DC}$ and $\delta a_{\rm orb, max}$ could be recovered independent of the rotation noise $\sigma_{\Delta\theta} \in [10^{-5}, 10^{-1}]^\circ$. 
The Eötvös parameter, however, is only recovered up to an uncertainty of $10^{-17}$ for rotation noise below a few $0.01^\circ$, which is well within reach of standard star-trackers and attitude control systems.

\section{Feasibility}\label{sec:feasibility}
In this section, we want to summarize and compute the previously derived requirements for the parameters in Tab.~\ref{tab:params}. The technical readiness level of a mission like STE-QUEST greatly benefits from the heritage of platform stability systems of previous missions. Thus, we also evaluate the requirements by analyzing the environment of previous and current missions, i.e., MICROSCOPE, LISA Pathfinder (LPF) and GRACE-FO\footnote{Note that, contrary to MICROSCOPE and LPF, GRACE-FO has no active drag-free control, and we only use the performance of the on board accelerometer \cite{christophe2015} as an estimate of the expected residual satellite accelerations.}.

Tab.~\ref{tab:feasibility} summarizes the constraints for STE-QUEST as well as the results for MICROSCOPE, LPF and GRACE-FO.
\begin{table*}[htb]
\caption{Requirements on satellite accelerations and rotations for $\eta = 10^{-15}$ and $\eta = 10^{-17}$. Both scenarios feature different atom interferometer sequences given in Tab.~\ref{tab:params}. The constraints are checked by analyzing the data from MICROSCOPE (\cite{pihan2019new} for acceleration and \cite{robert2022microscope} for rotation data), LISA Pathfinder (LPF) \cite{armano2019} and GRACE-FO \cite{christophe2015}. Note that $\Delta\Phi_i^a$ and $\Delta\delta\Phi_{\ddot{\theta}}$ are evaluated at frequencies $f T_c > 1$ whereas the quantities $\langle{\,\cdot{}\,}\rangle$ are constraints on smaller frequencies $f T_c < 1$.}
\begin{tabular}{c|c||c|c|c|c||c|c|c|c||c}
\hline \\ [-1.5ex]
Quantity & eq. & $\eta=10^{-15}$ & MICROSCOPE & LPF & GRACE-FO & $\eta=10^{-17}$ & MICROSCOPE & LPF & GRACE-FO & Unit \\
\hline 
\hline
$\Delta\Phi_i^a$ & \eqref{eq:acceleration-noise-req} & $\pi/10$ & $0.01$ & $0.11$  & $0.003$ & $\pi/10$ & $0.02$ & $0.22$  & $0.03$ & rad \\
$\langle\dot{a}\rangle$ & \eqref{eq:mean-linear-acc-req} & $3.2\times 10^{-11}$ & $1.0 \times 10^{-12}$ & $5.3 \times 10^{-11}$  & $2.0 \times 10^{-12}$ & $2.6\times 10^{-13}$ & $7.6 \times 10^{-14}$ & $5.4 \times 10^{-13}$  & $3.0 \times 10^{-13}$ & m/s$^3$ \\
$\Omega\vert_{\omega_{\rm orb}}$ & \eqref{eq:angular-velocity-sys-req} & $3.3 \times 10^{-9}$\footnote[1]{Can be relaxed by a factor $10^{3}$ to address the modulation of the EP-violating signal (cf. (ii) of Sec.~\ref{sec:Adv_Sat}).} & $1.2\times 10^{-8}$ & $4.0\times 10^{-10}$ & - & $3.3 \times 10^{-10}$\footnotemark[1] & $1.2\times 10^{-8}$ & $4.0\times 10^{-10}$ & - & rad/s \\
$\Omega\vert_{\omega_{\rm orb}/2}$ & \eqref{eq:angular-velocity-squared-sys-req} & $3.8 \times 10^{-5}$\footnotemark[1] & $3.7\times 10^{-8}$ & $3.8\times 10^{-10}$ & - &  $5.1 \times 10^{-6}$\footnotemark[1] & $3.7\times 10^{-8}$ & $3.8\times 10^{-10}$ & - & rad/s \\
$\Delta\delta\Phi_{\ddot{\theta}}$ & \eqref{eq:rotation-noise-req} & $4.5 \times 10^{-3}$ & $2.1 \times 10^{-5}$ & $1.9 \times 10^{-6}$  & - &  $8.9 \times 10^{-4}$ & $2.6 \times 10^{-5}$ & $1.5 \times 10^{-6}$  & - & rad \\
$\langle\Omega\rangle$ & \eqref{eq:mean-angular-vel-req} & $7.5 \times 10^{-6}$ & $1.3\times 10^{-9}$ & $2.9 \times 10^{-10}$ & - &  $5.4 \times 10^{-7}$ & $3.9\times 10^{-11}$ & $4.5 \times 10^{-12}$ & - & rad/s \\
$\langle\dot{\Omega}\rangle$ & \eqref{eq:mean-angular-acc-req} & $1.6 \times 10^{-5}$ & $4.1\times 10^{-9}$ & $8.8 \times 10^{-10}$ & - &  $1.3 \times 10^{-7}$ & $5.9\times 10^{-10}$ & $5.0 \times 10^{-11}$ & - & rad/s$^2$ \\
\hline
\end{tabular}
\label{tab:feasibility}
\end{table*}
For MICROSCOPE, we use the PSD of differential acceleration given in Ref.~\cite{pihan2019new}, as the satellite drag-free system is servo-controlled by one test mass, thus the differential acceleration between the test masses acts as an out of loop sensor for residual spacecraft accelerations. The rotation PSD is obtained from Ref.~\cite{robert2022microscope}, based on star-tracker data. For LPF and GRACE-FO, we use the PSDs presented in Ref.~\cite{armano2019} and Ref.~\cite{christophe2015}, respectively.
Integrating the PSDs together with the respective transfer functions (Eqs.~\eqref{eq:acceleration-noise-req}, \eqref{eq:rotation-noise-unc}) using a frequency band of $f\in [1/T_c, \infty)$ directly yields the uncertainty in the phase for a measurement of the differential acceleration. The quantities $\langle\,\cdot\,\rangle$ in Tab.~\ref{tab:feasibility}, that are requirements on the fluctuations in between cycles, are evaluated by integrating the respective PSD using frequencies smaller than the cycle frequency, i.e., $f\in (0,1/T_c]$. The angular velocity component modulated at orbital frequency, $\Omega\rvert_{\omega_{\rm orb}}$, is obtained by 
$\Omega\rvert_{\omega_{\rm orb}} = S_{\ddot{\theta}} (\omega_{\rm orb}) / \omega_{\rm orb}^2 / T_{\rm obs}$
where $S_{\ddot{\theta}}$ is the angular acceleration PSD that was obtained from a measurement with duration $T_{\rm obs}$. Analogously for $\Omega\rvert_{\omega_{\rm orb}/2}$.

In conclusion, most of the requirements are met with some margin, proving the technical readiness level of even a Q-WEP test at the $\eta = 10^{-17}$ level. 
For LPF, the change in the average linear acceleration in-between measurements, $\langle\dot{a}\rangle$, is about twice as large as the requirement. 
However, this is limited by out of loop noise on LPF, which could be reduced by acting on the laser frequency, which allows increasing the loop bandwidth. The same method can be applied to handle the slightly too large value of GRACE-FO.

\section{Conclusion}\label{sec:conclusion}

We have studied the platform requirements for a satellite-based dual-species atom interferometer testing the WEP beyond current state-of-the-art measurements.
In particular, we have derived the rotation, acceleration and orbit control requirements that a satellite needs to fulfill in order to allow a measurement of the Eötvös parameter $\eta$ to the unprecedented sensitivity of $10^{-17}$. We have demonstrated that the performance of previous (MICROSCOPE and LPF) and current (GRACE-FO accelerometer) satellite missions is sufficient to achieve the proposed sensitivity, underpinning the technical readiness of the STE-QUEST mission.

Additionally, we have derived requirements on self-gravity, temperature and magnetic field control inside the satellite at the payload location. To do so we have evaluated the effect of perturbing potentials up to order 4 of a polynomial expansion in position, which is beyond the reach of ``standard’’ methods (e.g. \cite{Storey1994}).

Ultimately, missions using atom interferometry are limited by atomic shot noise and the necessarily finite number of atoms that can be cooled and used. We anticipate that the development of entangled atomic source strategies ~\cite{Geiger18,Szigeti20,Corgier21,Corgier23} could reduce the constraints on the satellite platform and/or lead to better sensitivity to a violation of WEP.

\section{Acknowledgments}

The authors thank all contributors to STE-QUEST proposals (see Ref.~\cite{Ahlers2022} for a full list). R.C. thanks the Paris Observatory Scientific Council and was funded by ``PSL fellowship at Paris Observatory'' program. This work was funded by the Deutsche Forschungsgemeinschaft (German Research Foundation) under Germany’s Excellence Strategy (EXC-2123 QuantumFrontiers Grants No. 390837967) and through CRC 1227 (DQ-mat) within Projects No. A05, and the German Space Agency at the German Aerospace Center (Deutsche Raumfahrtagentur im Deutschen Zentrum f\"ur Luft- und Raumfahrt, DLR)  with funds provided by the German Federal Ministry of Economic Affairs and Climate Action due to an enactment of the German Bundestag under Grants Nos. 50WM2250A and 50WM2250E (QUANTUS+), No. 50WP1700 (BECCAL), No. 50WM2245A (CAL-II), No. 50WM2263A (CARIOQA-GE), No. 50WM2253A (AI-Quadrat), No. 50RK1957 (QGYRO), No. 50WM2177 (INTENTAS), as well as No. 50NA2106 (QGYRO+).

\appendix

\section{Phase-shift calculations}
\label{sec_App_A}
The phase shift $\Phi$ of a single atom interferometer arises from the overlap of two wave packets that travelled along different arms of the interferometer.
This propagation is encoded in the unitary time-evolution operator $\hat U_j$ associated with arm $j=1,2$ that ends in the considered exit port.
It includes the relevant momentum transfer at the points of atom-light interaction, as detailed below.
In a typical geometry, both arms are generated from the same initial wave packet $\ket{\psi}$, so that the expectation value of the overlap reads
\begin{equation}
    \mathcal{O }= \bra{\psi} \hat U_1^\dagger \hat U_2 \ket{\psi} = \mathcal{V} \exp (\ii \Phi),
\end{equation} 
where we have defined the visibility $\mathcal{V}$.
There are different methods to calculate the phase shift, such as in phase space~\cite{Giese2014,Dubetsky2016} or using path integrals \cite{Storey1994,Antoine2003,Overstreet2021,Overstreet2022}, as well as representation-free~\cite{Schleich2013,Kleinert2015} and perturbative techniques~\cite{ufrecht2020}.

\subsection{Path integrals}

The most commonly used method is based on path integrals~\cite{Storey1994}, where the overlap is usually obtained in position representation with $\psi( \xi)=\braket{ \xi | \psi}$ so that it takes (in one dimension) the form
\begin{equation}
\label{eq.overlap}
    \mathcal{O }= \iiint \dd x \dd \xi_1 \dd \xi_2 \psi^* ( \xi_1) \bra{\xi_1} \hat U_1^\dagger \ket{x} \bra{x } \hat U_2 \ket{ \xi_2}\psi ( \xi_2).
\end{equation} 
In principle, the propagator can be calculated by path integrals
\begin{equation}
     \bra{x} \hat U_j \ket{ \xi_j} = \int\limits_{ \xi_j}^x \mathcal{D} \mathcal{X} \exp (\ii S_j[  \mathcal{X}]/\hbar),
\end{equation}
where the action functional $S_j[ \mathcal{X}]$ depends on the Hamiltonian that describes the motion along arm $j$, so that Eq.~\eqref{eq.overlap} corresponds to the influence functional~\cite{Feynman1965,Su1988}.
For an exact description the complete path integral has to be evaluated, and so far there is no connection to a classical trajectory, let alone the classical action, since the evaluation requires an integration over \textit{all} trajectories $ \mathcal{X}$.
Hence, the path integral is in general not associated with any vanishing variation of the action~\cite{DeWitt2003}.

However, one can write $ \mathcal{X}= x_j + \nu$, where $x_j(t)$ is the classical trajectory with initial condition $\xi_j$ and $\nu (t)$ are fluctuations with vanishing initial conditions.
Expanding the action around the classical trajectories $x_j$, the linear order of expansion vanishes since the classical trajectory follows Euler-Lagrange equations.
Hence, one arrives~\cite{Feldbrugge2023} at
\begin{equation}
     \bra{x} \hat U_j \ket{\xi_j} = \exp (\ii S_j[x_j ]/\hbar) \int\mathcal{D} \nu  \exp (\ii \delta S_j[\nu]/\hbar)
\end{equation}
with 
\begin{equation}
\delta S_j[\nu ] =\frac{\partial^2 S_j}{2\partial x_j^2} \nu^2 + \frac{\partial^2 S_j}{2\partial \dot x_j^2 } \dot \nu^2 + \frac{\partial^2 S_j}{2\partial x_j \partial \dot x_j}  \nu \dot \nu + \cdots ,
\end{equation}
where the derivatives are evaluated at $  \nu =0$ and the remaining path integral is the fluctuation integral.
It can be calculated exactly for linear potentials and is independent of initial and final conditions.
Since the momentum transfer that acts during beam splitter and mirror pulses can be modeled by a linear potential, a calculation for simple cases of atom interferometers is straightforward.

For a symmetric double-diffraction Mach-Zehnder interferometer with interrogation time $T$ and acceleration $a$ that closes in phase space, one arrives at a phase
\begin{equation}
\label{eq.action-difference}
    (S_2-S_1)/\hbar = 2 k a T^2 + \Phi_{\rm L }(-T) - 2 \Phi_{\rm L}(0) + \Phi_{\rm L}(T)
\end{equation}
and no further phase contributions arise from the fluctuation integral.
In this case, one can associate the phase with the difference of classical action $ S_2-S_1 $.
Here, $\Phi_{\rm L} (t)$ is the laser phase at time $t$ and its discrete second derivative enters the phase. 

Moreover, the fluctuation integral can also be evaluated for general quadratic Lagrangians, and hence, as an approximation for higher orders using the method of stationary phase.
However, in this case the observed phase difference depends not solely on the difference of actions, because the wave packets do not generally overlap perfectly in phase space~\cite{Roura2014} and a separation phase occurs~\cite{Overstreet2021}.
In addition, depending on the particular potential, the shape of the wave packets might change and lead to additional contributions.
Nevertheless, one can use path integrals to calculate perturbative effects of rotations and quadratic potentials~\cite{Storey1994}.

Such techniques can be used for phase estimations that are based on classical actions, also for potentials beyond quadratic order and without explicitly evaluating the fluctuation integral~\cite{Overstreet2022}.
While these approaches give some insight into possible phase contributions, they are effectively semiclassical.

When evaluating the difference of actions between both classical trajectories, the virial theorem can be used to express the corresponding integrals solely through the midpoint trajectory $(x_1+ x_2)/2$ and the arm separation $x_1 - x_2$ for a linear potential~\cite{Loriani2019}.
One can also perform an expansion of the potential around this midpoint trajectory in orders of the arm separation~\cite{Overstreet2021}.
This way, one obtains a convenient tool for a calculation of the action difference, but not for obtaining the exact Feynman propagator for arbitrary potentials by evaluating the fluctuation integral.
One can circumvent such problems by treating weak potentials as a perturbation.

\subsection{Perturbative operator method}

We rely on a perturbative but operator-valued method following the work of \cite{ufrecht2020}.
Let us assume, that the Hamiltonian inducing the motion along arm $j$ takes the form $\hat{H}_j= \hat{\mathcal{H}}_j + \hat{V}$.
The first, unperturbed contribution
\begin{equation*}
    \hat{\mathcal{H}}_j= \frac{\hat p^2}{2 m} - \hbar \sum_\ell \big[k_j^{(\ell)} \hat x + \Phi_\text{L}^{(\ell,j)} (t) \big] \delta (t-T_\ell)
\end{equation*}
includes the momentum transfer $k_j^{(\ell)}$ on arm $j$ of pulse $\ell$ at time $T_\ell$ (being equal to $-T$, $0$ and $T$ respectively at the first, second and third pulse), and we have included the corresponding laser phase $\Phi_\text{L}^{(\ell,j)}$ in this effective potential.
Moreover, we assume that the perturbing potential has the form
\begin{equation}
         V(\hat x)= \sum_{n=1}^N c_{n} \hat x^n
         \label{eq:potential-series-expansion-appendix}
\end{equation}
that also incorporates a linear term, which accounts for perturbative accelerations and, in particular, accelerations in microgravity.

We then change into the interaction picture~\cite{ufrecht2020} with respect to the unperturbed Hamiltonian $ \hat{\mathcal{H}}_j $ to calculate the overlap from Eq.~\eqref{eq.overlap}.
We make use of fact that the phase for a closed~\cite{Roura2014}, unperturbed atom interferometer can be trivially calculated, e.\,g., using path integrals~\cite{Storey1994} as described above or by an representation-free method~\cite{Schleich2013}.
The remaining part of the overlap amounts to perturbatively calculating the Schwinger-Keldysh closed-time-path Green's function, which is equivalent to evaluating the influence functional known from the path-integral formalism~\cite{Su1988}.
Using a combination of Magnus expansion and cumulant expansion, one can show~\cite{ufrecht2020} that $\Phi = \Phi_{\rm L }(-T) - 2 \Phi_{\rm L}(0) + \Phi_{\rm L}(T) + \Phi_\text{pert}$, directly obtained from the overlap, consists of a phase induced by the unperturbed Hamiltonian and additional perturbations $ \Phi_\text{pert}$ that can be divided into two contributions, namely
\begin{equation}
\begin{split}
    \Phi_{\rm pert} =& - \frac{1}{\hbar} \int\limits_{-T}^{T} dt\, \left[ V(x_{1}(t)) - V(x_{2}(t))\right] \\
    &- \frac{1}{2\hbar}  \int\limits_{-T}^{T} dt\, \left[ \frac{\partial^2 V}{\partial^2 x} \Bigg|_{x_1(t)}- \frac{\partial^2 V}{\partial^2 x}\Bigg|_{x_2(t)} \right] \left[\sigma_{x}^2 + \sigma_{v}^2 t^2\right].
    \label{eq:simplified-phase-giese-appendix}
\end{split}
\end{equation}
Here, $x_j(t)$ describes the classical unperturbed trajectory along arm $j$, i.\,e., the one induced by the classical analog of $\hat{\mathcal H}_j$.
Moreover, $\sigma_{x}^2$ is the initial width in position of the wave packet and $\sigma_{v}^2$ is its initial velocity width.
For this form, we have assumed that there is initially no correlation between position and momentum.
The first contribution is just the perturbing potential $V$ evaluated at the classical, unperturbed trajectories of both arms, whereas the second contribution accounts for imperfect overlap of wave packets due to their deformation.

We find for harmonic, cubic, and quartic perturbations, i.\,e., for $N=4$, in a double-diffraction Mach-Zehnder interferometer
\begin{equation}
\begin{split}
    \Phi_{\rm pert} = &- \frac{2k T^2}{m}c_{1}  -2\frac {2k T^2}{m}x(T)c_{2} + \kappa c_3\\
&+ 4\left[\kappa x(T) + \dfrac{k T^2}{m}(4x^3(T)-2T^3 v_{0} \sigma_{v}^2)\right]c_{4},
\end{split}
\end{equation}
with the abbreviation
\begin{equation*}
    \kappa=- \dfrac{k T^2}{m}\left[6x^2(T)+6\sigma^2_{x}+T^2\left(v^2_{0}+\left(\dfrac{\hbar k}{m}\right)^2+ 7\sigma^2_{v}\right)\right]
\end{equation*}
and $x(T)=x_0 + v_0 T$, where $x_0$ and $v_0$ correspond to the initial expectation value of position and velocity, respectively.
In particular, we observe that wave packet deformations arise for cubic potentials, whereas for quadratic potentials initial conditions enter because the perturbation causes the interferometer to open~\cite{Roura2014}.
The first term that stems from linear potentials has exactly the form $2 k a T^2 $ discussed in Eq.~\eqref{eq.action-difference}.

\subsection{Deriving constraints on the potential's coefficients}
In order to reach the proposed target uncertainty, we require that each spurious phase shift induced by the different coefficients of the potential in Eq.~\eqref{eq:potential-series-expansion-appendix} is smaller than the Eötvös signal $2\eta g_0 k_\mathrm{AB} T^2$. It should be noted that the coefficients $c_{n}$ scale different parameters in the differential phase:
\begin{equation}
\begin{split}
    \delta\Phi = &\delta\Phi_1 (k_i,T,m_i) c_1 + \delta\Phi_2 (k_i,T,m_i,x_{0,i},v_{0,i}) c_2 \\
    &+ \sum_{j=3}^N \delta\Phi_j (k_i,T,m_i,x_{0,i},v_{0,i}, \sigma_{x,i}, \sigma_{v,i}) c_j ,
\end{split}
\end{equation}
where $i$ marks the species and assuming $c_n$ is species independent. The terms that are independent of statistical parameters can be suppressed by knowing the value of $c_n$ up to an uncertainty $\Delta c_n$ leaving:
\begin{equation}
\begin{split}
    \Delta\delta\phi = &|\delta\Phi_1| \Delta c_1 + \sum_{j=2}^N [ |\Delta\delta\Phi_j (\Delta \delta x_{0},\Delta \delta v_{0})| c_j \\
    & + |\delta\Phi_j ( \delta x_{0}, \delta v_{0})| \Delta c_j + |\Delta\delta\Phi_j (\Delta \delta x_{0},\Delta \delta v_{0})|\Delta c_j]
\end{split}
\end{equation}
where $\Delta\delta\Phi_j (\Delta\delta x_0, \Delta\delta v_0) = |\partial\delta\Phi_j / (\partial \delta x_0)| \Delta\delta x_0 + |\partial\delta\Phi_j / (\partial \delta v_0)| \Delta\delta v_0$.
Note that this is a pessimistic treatment. Since most of these contributions are uncorrelated, the favorable quadratic sum would also suffice.

Thus, we get requirements on the nominal values $c_n$ for $n>2$ and on the uncertainties $\Delta c_n$ for all $n$:
\begin{equation}
\begin{split}
    c_j &\leq 2 \eta g k_\mathrm{AB} T^2 / |\Delta\delta\Phi_j| , \quad j \in \mathbb{N} \setminus \{1  \} \\
    \Delta c_j &\leq 2 \eta g k_\mathrm{AB} T^2 / (|\delta\Phi_j| + |\Delta\delta\Phi_j| ) , \quad j \in \mathbb{N}.
\end{split}
\end{equation}

For some applications, the potential might not only be linear in the coefficients of interest. For example, for the black body radiation potential we find
\begin{equation}
    V_{\rm BBR}(x)=\dfrac{2\alpha_i \sigma }{c \epsilon _0} T_{\rm tube}^4(x), \quad T_{\rm tube}(x) = \sum_{n=0} t_n x^n
\end{equation}
yielding $c_n = f(\{ t_{m}\}_{m=0}^n)$ where $f$ denotes an arbitrary function. Thus, the constraints on $t_n$ are codependent where the constraint on $t_n$ depends on the value of $t_{n-1}$,
\begin{equation}
\begin{split}
    t_j &\leq 2 \eta g k_\mathrm{AB} T^2 / \Delta\delta\Phi_j (\{ t_{m}\}_{m=0}^{j-1}) , \quad j \in \mathbb{N} \setminus \{1  \}, \\
    \Delta t_j &\leq 2 \eta g k_\mathrm{AB} T^2 / (|\partial_{t_j}\delta\Phi (\{ t_{m}\}_{m=0}^n)| \\
    & \phantom{\leq 2 \eta g k_\mathrm{AB} T^2 / (} + |\partial_{t_j}\Delta\delta\Phi (\{ t_{m}\}_{m=0}^n)|) ,\quad j \in \mathbb{N}.
\end{split}
\end{equation}
This set of equations can be solved by assuming values for $t_0$, $\Delta t_0$ and $t_1$. The second-order Zeeman effect can be treated analogously.

\section{Deriving the angular acceleration transfer function}\label{sec:rotation-derivation}

For the following derivation of the angular acceleration transfer function, we look at the configuration shown in Fig.~\ref{fig:rotation-setup} and work in an inertial frame whose origin coincides with the center of mass of the satellite for all times $t$. 
The orientation is chosen such that the sensitive axis of the interferometer is aligned with the $x$-axis when the atoms are initialized.
The mirror is assumed to be rectangular with a thickness of $d_{\rm M}$ with its center of mass positioned at $\bf{r}_{\rm M}$.
Small rotations of the mirror $\theta_{\rm M}$ simply add to the effect of the rotation of the satellite $\theta_{\rm S}$ such as the overall rotation is $\theta(t) = \theta_{\rm M}(t) + \theta_{\rm S}(t)$.

In the inertial frame, the effective wave vector reads ${\bf{k}}_i = - k_i \times (\cos [\theta(t)], \sin[\theta(t)])^T$ and the initial velocity of the atoms is defined as ${\bf{v}}_0 = \tilde{\bf{v}}_0 + {\bf{\Omega}} \times \tilde{\bf{r}}_0$. 
For simplicity, initial accelerations are neglected. 
Here, $\tilde{\bf{r}}_0$ and $\tilde{\bf{v}}_0$ denote the initial position and velocity of the atoms in the rotating satellite frame.
The resulting interferometer phase, $\Phi_i$, can be deduced from the effective laser phase $\phi_{i} (t) = {\bf{k}}_{i}(t) \cdot {\bf{r}}_{i}(t)$ where ${\bf{k}}_{i}(t)$ and ${\bf{r}}_{i}(t)$ are respectively the effective wave-vector and the c.m. position of species $i$ at pulse at time $t \in \{-T,0,T\}$.  
We find

\begin{equation}
\begin{split}
    \phi_i(t) =& k_i (\tilde r_{x,i} (T_0+T+t) - (r_{\text{M},x} + d_\text{M})) \\
    &+ k_i \tilde r_{y,i} (T_0+T+t) \theta(t) - \tfrac{1}{2}k_i \tilde r_{x,i} (T_0+T+t) \theta (t)^2 \\
    & + \mathcal O (\theta(t)^3),
\end{split}
\end{equation}
where $\tilde r_{x,i}(t) = r_{x,0,i} + (v_{x,0,i} - r_{y,0,i} \Omega_0) t$ and $\tilde r_{y,i}(t) = r_{y,0,i} + (v_{y,0,i} + r_{x,0,i} \Omega_0) t$ denote the atoms' position in the rotated reference frame and $T_0$ is the dead time measuring the duration from the release of the atoms until the first laser pulse at $t=-T$ is applied. $\Omega_0 = d\theta/dt\vert_{t=-T_0-T}$ denotes the angular velocity of the satellite at the release of the atoms.
In the final atom interferometer phase, $\Phi_i = 2[\phi_{i}(-T) - 2\phi_{i}(0) + \phi_{i}(T)]$, all terms that are constant or linear in time vanish:
\begin{equation}
\begin{split}
    \Phi_i = &2 k_i [
        \tilde{r}_{y,i}(T_0) \theta (-T) - 2 \tilde{r}_{y,i}(T+T_0)\theta(0) \\
        &\phantom{2 k_i [} + \tilde{r}_{y,i}(T_0 + 2T)\theta(T)] \\ 
        & + \mathcal O (\theta(t)^2).
\end{split}
\label{eq:rotations-single-ati-phase}
\end{equation}

The sensitivity function of the differential phase $\delta\Phi$ (see Eq.~\eqref{eq_AI_phase2}) for a jump in the satellite rotation $\Delta \theta (t)$ at time $t$ is defined as \cite{cheinet2008}
\begin{equation}
    g_\theta(t) = \lim_{\Delta \theta \rightarrow 0} \frac{\Delta \delta\Phi(\Delta\theta(t))}{\Delta\theta(t)}.
\end{equation}
Inserting Eq.~\eqref{eq:rotations-single-ati-phase} yields
\begin{equation}
    g_\theta (t) = 2k_{AB}\times \begin{cases}
        0, & t<-T \\
        -(\delta r_{y,0} + \delta v_{y,0} T_0), & -T<t<0 \\
        (\delta r_{y,0} + \delta v_{y,0} (T_0+2T)), & 0<t<T \\
        0, & t>T
    \end{cases},
\end{equation}
where $k_{AB} = 2k_Ak_B/(k_A+k_B)$.
With this, we immediately obtain the corresponding transfer function by taking the Fourier transform of the sensitivity function \cite{cheinet2008}:
\begin{equation}
\begin{split}
    H_\theta (\omega) 
    = 4 k_{AB} [
    & -2 i (\delta r_{y,0} + \delta v_{y,0} (T_0 + T) \sin^2 (\omega T /2) \\
    & + \delta v_{y,0} T \sin (T \omega) ].
    \label{eq:rot-transfer-func-appendix}
\end{split}
\end{equation}

\section{Analyzing systematic uncertainties using the satellite simulator}\label{appendix:squid}
In this section, we go into more detail how the SQUID simulator analyzes systematic and statistical uncertainties, using the orbit analysis in Sec.~\ref{sec:orbit-control} as an example.
The signal under consideration is given by
\begin{equation}
    \delta\Phi (r_S) = 2 [\eta g_x(r_S) + 2\delta r_{x,0} \Gamma_{xx}(r_S) ] k_{AB} T^2 ,
    \label{eq:ATI-signal-analysis-appendix}
\end{equation}
where $r_S$ denotes the satellite's position and $\Gamma_{xx}$ is the $xx$-component of the gravity gradient. For simplicity, we focus only on $\Gamma_{xx}$, but this can easily extended to include the other components. Note that in the following the signal is sampled at certain satellite positions $r_S(t_i)$ where $t_i$ marks the times a measurement is performed: $t_{i+1} = t_i + T_c$.

For the numerical analysis, we construct the model used for the fit according to
\begin{equation}
    \mathcal M = 2\begin{pmatrix}
        g_x(r_S(t_0)) & \Gamma_{xx}(r_S(t_0)) \\
        \cdots & \cdots \\
        g_x(r_S(t_{n-1})) & \Gamma_{xx}(r_S(t_{n-1}))
    \end{pmatrix} k T^2 ,
    \label{eq:ATI-model-analysis}
\end{equation}
for the free parameters $p_f = (\eta, \delta r_{x,0} )^T$ such that $\delta\Phi = (\delta\Phi(r_S(t_0)),\dots ,\delta\Phi(r_S(t_{n-1})))^T =  \mathcal M \cdot p_f$.

Our analysis is performed in the following steps
\begin{enumerate}
    \item Generate a signal $\delta\tilde\Phi (r_S + \Delta r_S)$ assuming a certain value for $\eta$ and $\delta r_{x,0}$ (see Eq.~\eqref{eq:ATI-signal-analysis-appendix}) for a distorted circular orbit according to the Hill model $\Delta r_S = (\Delta R$, $\Delta T$, $\Delta N)$ (see Eq.~\eqref{eq:hill-model}) that is additionally subject to white noise (i.e., atomic shot noise).
    \item Generate a model matrix $\mathcal M$ according to Eq.~\eqref{eq:ATI-signal-analysis-appendix} using the undistorted circular orbit from (1.).
    \item Fit the generated signal (1.) using the model (2.). for the free parameters $\eta$ and $\delta r_{x,0}$ for various distortion strengths $\Delta r_S$.
\end{enumerate}

The fit can be obtained by a Generalized Least Squares (GLS) analysis where the best possible estimate of the free parameters $p_f$ is obtained by
\begin{equation}
    p_f^{\rm GLS} = (\mathcal M^T \Omega^{-1} \mathcal M )^{-1} \mathcal M^T \Omega^{-1} \delta\tilde\Phi
\end{equation}
where $\Omega$ is the covariance matrix and $v_{\rm GLS} = (\mathcal M ^T \Omega^{-1} \mathcal M)^{-1}$ the variance-covariance matrix. Assuming a white noise model, $\Omega$ reduces to $\Omega = \sigma \mathbf{1}$ where $\sigma$ is the width of the distribution.
In this case, the analysis simplifies to the Ordinary Least Squares (OLS) method defined as
\begin{equation}
    p_f^{\rm OLS} = (\mathcal M^T \mathcal M )^{-1} \mathcal M^T \delta\tilde\Phi
\end{equation}
with its variance-covariance matrix being $v_{\rm OLS} = \sigma^2 (\mathcal M ^T \mathcal M)^{-1}$.

\bibliographystyle{unsrt}
\bibliography{references}

\begin{thebibliography}{10}

\bibitem{Kiefer}
C.~Kiefer.
\newblock {\em Quantum gravity}.
\newblock Oxford University Press, Oxford., 2007.

\bibitem{Will2018}
Clifford.~M. Will.
\newblock {\em Theory and Experiment in Gravitational Physics}.
\newblock Cambridge University Press, 2018.

\bibitem{10.1119/1.4895342}
Eolo Di~Casola, Stefano Liberati, and Sebastiano Sonego.
\newblock {Nonequivalence of equivalence principles}.
\newblock {\em American Journal of Physics}, 83(1):39--46, 01 2015.

\bibitem{damour:1994fk}
T.~{Damour} and A.~M. {Polyakov}.
\newblock {The string dilation and a least coupling principle}.
\newblock {\em Nuclear Physics B}, 423:532--558, July 1994.

\bibitem{fayet:1974ww}
P.~{Fayet} and J.~{Iliopoulos}.
\newblock {Spontaneously broken supergauge symmetries and goldstone spinors}.
\newblock {\em Physics Letters B}, 51(5):461--464, September 1974.

\bibitem{fayet:1990tu}
Pierre {Fayet}.
\newblock {Extra U(1)'s and new forces}.
\newblock {\em Nuclear Physics B}, 347(3):743--768, December 1990.

\bibitem{Fayet:2017pdp}
Pierre Fayet.
\newblock {MICROSCOPE limits for new long-range forces and implications for
  unified theories}.
\newblock {\em Phys. Rev. D}, 97(5):055039, 2018.

\bibitem{Kostelecky:1994rn}
V.~Alan Kostelecky and Robertus Potting.
\newblock {CPT, strings, and meson factories}.
\newblock {\em Phys. Rev. D}, 51:3923--3935, 1995.

\bibitem{damour:2002ys}
T.~{Damour}, F.~{Piazza}, and G.~{Veneziano}.
\newblock {Runaway Dilaton and Equivalence Principle Violations}.
\newblock {\em Physical Review Letters}, 89(8):081601, August 2002.

\bibitem{damour:2010zr}
Thibault Damour and John~F. Donoghue.
\newblock Equivalence principle violations and couplings of a light dilaton.
\newblock {\em Phys. Rev. D}, 82:084033, Oct 2010.

\bibitem{fayet:2019aa}
P.~{Fayet}.
\newblock {M I C R O S C O P E limits on the strength of a new force with
  comparisons to gravity and electromagnetism}.
\newblock {\em Physical Review D}, 99(5):055043, March 2019.

\bibitem{Touboul2022}
Pierre Touboul, Gilles M\'etris, Manuel Rodrigues, Joel Berg\'e, Alain Robert,
  Quentin Baghi, Yves Andr\'e, Judica\"el Bedouet, Damien Boulanger, Stefanie
  Bremer, Patrice Carle, Ratana Chhun, Bruno Christophe, Valerio Cipolla,
  Thibault Damour, Pascale Danto, Louis Demange, Hansjoerg Dittus, Oc\'eane
  Dhuicque, Pierre Fayet, Bernard Foulon, Pierre-Yves Guidotti, Daniel
  Hagedorn, Emilie Hardy, Phuong-Anh Huynh, Patrick Kayser, St\'ephanie Lala,
  Claus L\"ammerzahl, Vincent Lebat, Fran\ifmmode
  \mbox{\c{c}}\else~\c{c}\fi{}oise Liorzou, Meike List, Frank L\"offler,
  Isabelle Panet, Martin Pernot-Borr\`as, Laurent Perraud, Sandrine Pires,
  Benjamin Pouilloux, Pascal Prieur, Alexandre Rebray, Serge Reynaud, Benny
  Rievers, Hanns Selig, Laura Serron, Timothy Sumner, Nicolas Tanguy, Patrizia
  Torresi, and Pieter Visser.
\newblock Microscope mission: Final results of the test of the equivalence
  principle.
\newblock {\em Phys. Rev. Lett.}, 129:121102, Sep 2022.

\bibitem{Wagner2012}
T~A Wagner, S~Schlamminger, J~H Gundlach, and E~G Adelberger.
\newblock Torsion-balance tests of the weak equivalence principle.
\newblock {\em Classical and Quantum Gravity}, 29(18):184002, 2012.

\bibitem{Williams09}
JAMES~G. WILLIAMS, SLAVA~G. TURYSHEV, and DALE~H. BOGGS.
\newblock Lunar laser ranging tests of the equivalence principle with the earth
  and moon.
\newblock {\em International Journal of Modern Physics D}, 18(07):1129–1175,
  Jul 2009.

\bibitem{jonsson1961elektroneninterferenzen}
Claus J{\"o}nsson.
\newblock Elektroneninterferenzen an mehreren k{\"u}nstlich hergestellten
  feinspalten.
\newblock {\em Zeitschrift f{\"u}r Physik}, 161(4):454--474, 1961.

\bibitem{gerlach1922experimentelle}
Walther Gerlach and Otto Stern.
\newblock Der experimentelle nachweis der richtungsquantelung im magnetfeld.
\newblock {\em Zeitschrift f{\"u}r Physik}, 9(1):349--352, 1922.

\bibitem{Parker18}
R.H. Parker, W~Zhong C.~Yu, B.~Estey, and H.~Müller.
\newblock Measurement of the fine-structure constant as a test of the standard
  model.
\newblock {\em Nature}, 360:191--195, 2018.

\bibitem{Morel20}
L.~Morel, Z.~Yao, P.~Cladé, and S.~Guellati-Khélifa.
\newblock Determination of the fine-structure constant with an accuracy of 81
  parts per trillion.
\newblock {\em Nature}, 588:61–65, Dec 2020.

\bibitem{Sorrentino14}
F.~Sorrentino, Q.~Bodart, L.~Cacciapuoti, Y.-H. Lien, M.~Prevedelli, G.~Rosi,
  L.~Salvi, and G.~M. Tino.
\newblock Sensitivity limits of a raman atom interferometer as a gravity
  gradiometer.
\newblock {\em Phys. Rev. A}, 89:023607, Feb 2014.

\bibitem{Rosi14}
G.~Rosi, F.~Sorrentino, L.~Cacciapuoti, M.~Prevedelli, and G.~M. Tino.
\newblock Precision measurement of the newtonian gravitational constant using
  cold atoms.
\newblock {\em Nature}, 510:518–521, 2014.

\bibitem{Geiger20}
R.~Geiger, A.~Landragin, S.~Merlet, and F.~Pereira~Dos Santos.
\newblock High-accuracy inertial measurements with cold-atom sensors.
\newblock {\em AVS Quantum Sci.}, 1, Jun 2020.

\bibitem{Canuel2017}
B.~{Canuel}, A.~{Bertoldi}, L.~{Amand}, E.~{Borgo di Pozzo}, B.~{Fang},
  R.~{Geiger}, J.~{Gillot}, S.~{Henry}, J.~{Hinderer}, D.~{Holleville},
  G.~{Lef{\`e}vre}, M.~{Merzougui}, N.~{Mielec}, T.~{Monfret}, S.~{Pelisson},
  M.~{Prevedelli}, S.~{Reynaud}, I.~{Riou}, Y.~{Rogister}, S.~{Rosat},
  E.~{Cormier}, A.~{Landragin}, W.~{Chaibi}, S.~{Gaffet}, and P.~{Bouyer}.
\newblock {Exploring gravity with the MIGA large scale atom interferometer}.
\newblock {\em Scientific Reports}, 8:14064, 2018.

\bibitem{Canuel20}
B.~Canuel, S.~Abend, P.~Amaro-Seoane, F.~Badaracco, Q.~Beaufils, A.~Bertoldi,
  K.~Bongs, P.~Bouyer, C.~Braxmaier, W.~Chaibi, N.~Christensen, F.~Fitzek,
  G.~Flouris and. N.~Gaaloul, S.~Gaffet, C.~L.~Garrido Alzar, R.~Geiger,
  S.~Guellati-Khelifa, K.~Hammerer, J.~Harms, J.~Hinderer, M.~Holynski,
  J.~Junca, S.~Katsanevas, C.~Klempt, C.~Kozanitis, M.~Krutzik, A.~Landragin,
  I.~Lazaro Roche, B.~Leykauf, Y-H. Lien, S.~Loriani, S.~Merlet, M.~Merzougui,
  M.~Nofrarias, P.~Papadakos, F.~Pereira dos Santos, A.~Peters, D.~Plexousakis,
  M.~Prevedelli, E.~M. Rasel, Y.~Rogister, S.~Rosat, A.~Roura, D.~O. Sabulsky,
  V.~Schkolnik, D.~Schlippert, C.~Schubert, L.~Sidorenkov, J-N. Siemß, C.~F.
  Sopuerta, F.~Sorrentino, C.~Struckmann, G.~M. Tino, G.~Tsagkatakis,
  A.~Vicere, W.~von Klitzing, L.~Woerner, and X~Zou.
\newblock Elgar—a european laboratory for gravitation and
  atom-interferometric research.
\newblock {\em Class. Quantum Grav.}, 37(22), 2020.

\bibitem{Badurina20}
L.~Badurina, E.~Bentine, D.~Blas, K.~Bongs, D.~Bortoletto, T.~Bowcock,
  K.~Bridges, W.~Bowden, O.~Buchmueller, C.~Burrage, J.~Coleman, G.~Elertas,
  J.~Ellis, C.~Foot, V.~Gibson, M.G. Haehnelt, T.~Harte, S.~Hedges, R.~Hobson,
  M.~Holynski, T.~Jones, M.~Langlois, S.~Lellouch, M.~Lewicki, R.~Maiolino,
  P.~Majewski, S.~Malik, J.~March-Russell, C.~McCabe, D.~Newbold, B.~Sauer,
  U.~Schneider, I.~Shipsey, Y.~Singh, M.A. Uchida, T.~Valenzuela, M.~van~der
  Grinten, V.~Vaskonen, J.~Vossebeld, D.~Weatherill, and I.~Wilmut.
\newblock Aion: an atom interferometer observatory and network.
\newblock {\em Journal of Cosmology and Astroparticle Physics}, 2020(05):011,
  may 2020.

\bibitem{Abe2021}
Mahiro Abe, Philip Adamson, Marcel Borcean, Daniela Bortoletto, Kieran Bridges,
  Samuel~P Carman, Swapan Chattopadhyay, Jonathon Coleman, Noah~M Curfman,
  Kenneth DeRose, Tejas Deshpande, Savas Dimopoulos, Christopher~J Foot,
  Josef~C Frisch, Benjamin~E Garber, Steve Geer, Valerie Gibson, Jonah Glick,
  Peter~W Graham, Steve~R Hahn, Roni Harnik, Leonie Hawkins, Sam Hindley,
  Jason~M Hogan, Yijun Jiang, Mark~A Kasevich, Ronald~J Kellett, Mandy Kiburg,
  Tim Kovachy, Joseph~D Lykken, John March-Russell, Jeremiah Mitchell, Martin
  Murphy, Megan Nantel, Lucy~E Nobrega, Robert~K Plunkett, Surjeet Rajendran,
  Jan Rudolph, Natasha Sachdeva, Murtaza Safdari, James~K Santucci, Ariel~G
  Schwartzman, Ian Shipsey, Hunter Swan, Linda~R Valerio, Arvydas Vasonis,
  Yiping Wang, and Thomas Wilkason.
\newblock Matter-wave atomic gradiometer interferometric sensor ({MAGIS}-100).
\newblock {\em Quantum Science and Technology}, 6(4):044003, jul 2021.

\bibitem{Zhan2020}
Ming-Sheng Zhan, Jin Wang, Wei-Tou Ni, Dong-Feng Gao, Gang Wang, Ling-Xiang He,
  Run-Bing Li, Lin Zhou, Xi~Chen, Jia-Qi Zhong, Biao Tang, Zhan-Wei Yao, Lei
  Zhu, Zong-Yuan Xiong, Si-Bin Lu, Geng-Hua Yu, Qun-Feng Cheng, Min Liu,
  Yu-Rong Liang, Peng Xu, Xiao-Dong He, Min Ke, Zheng Tan, and Jun Luo.
\newblock Zaiga: Zhaoshan long-baseline atom interferometer gravitation
  antenna.
\newblock {\em International Journal of Modern Physics D}, 29(04):1940005,
  2020.

\bibitem{ArxivCERN}
Arduini G., Badurina L., Balazs K., Baynham C., Buchmueller O., Buzio M.,
  Calatroni S., Corso J.-P., Ellis J., Gaignant Ch., Guinchard M., Hakulinen
  T., Hobson R., Infantino A., Lafarge D., Langlois R., Marcel C., Mitchell J.,
  Parodi M., Pentella M., Valuch D., and Vincke H.
\newblock A long-baseline atom interferometer at cern: Conceptual feasibility
  studyd gravitational wave detection.
\newblock {\em arXiv preprint arXiv:2304.00614}, 2023.

\bibitem{PhysRevLett.112.203002}
D.~Schlippert, J.~Hartwig, H.~Albers, L.~L. Richardson, C.~Schubert, A.~Roura,
  W.~P. Schleich, W.~Ertmer, and E.~M. Rasel.
\newblock Quantum test of the universality of free fall.
\newblock {\em Phys. Rev. Lett.}, 112:203002, May 2014.

\bibitem{PhysRevLett.115.013004}
Lin Zhou, Shitong Long, Biao Tang, Xi~Chen, Fen Gao, Wencui Peng, Weitao Duan,
  Jiaqi Zhong, Zongyuan Xiong, Jin Wang, Yuanzhong Zhang, and Mingsheng Zhan.
\newblock Test of equivalence principle at $1{0}^{\ensuremath{-}8}$ level by a
  dual-species double-diffraction raman atom interferometer.
\newblock {\em Phys. Rev. Lett.}, 115:013004, Jul 2015.

\bibitem{Asenbaum2020}
Peter Asenbaum, Chris Overstreet, Minjeong Kim, Joseph Curti, and Mark~A.
  Kasevich.
\newblock Atom-interferometric test of the equivalence principle at the
  ${10}^{\ensuremath{-}12}$ level.
\newblock {\em Phys. Rev. Lett.}, 125:191101, Nov 2020.

\bibitem{Zhang_2020}
Ke~Zhang, Min-Kang Zhou, Yuan Cheng, Le-Le Chen, Qin Luo, Wen-Jie Xu, Lu-Shuai
  Cao, Xiao-Chun Duan, and Zhong-Kun Hu.
\newblock Testing the universality of free fall by comparing the atoms in
  different hyperfine states with bragg diffraction*.
\newblock {\em Chinese Physics Letters}, 37(4):043701, apr 2020.

\bibitem{10.1116/5.0076502}
B.~Barrett, G.~Condon, L.~Chichet, L.~Antoni-Micollier, R.~Arguel, M.~Rabault,
  C.~Pelluet, V.~Jarlaud, A.~Landragin, P.~Bouyer, and B.~Battelier.
\newblock {Testing the universality of free fall using correlated 39K–87Rb
  atom interferometers}.
\newblock {\em AVS Quantum Science}, 4(1):014401, 02 2022.

\bibitem{Kovachy2016}
Tim Kovachy, Jason~M. Hogan, Alex Sugarbaker, Susannah~M. Dickerson,
  Christine~A. Donnelly, Chris Overstreet, and Mark~A. Kasevich.
\newblock Matter wave lensing to picokelvin temperatures.
\newblock {\em Phys. Rev. Lett.}, 114:143004, Apr 2015.

\bibitem{Corgier18}
R.~Corgier, S.~Amri, W.~Herr, H.~Ahlers, J.~Rudolph, D.~Guery-Odelin, E.~M.
  Rasel, E.~Charron, and N.~Gaaloul.
\newblock Fast manipulation of bose-einstein condensates with an atom chip.
\newblock {\em New Journal of Physics}, 20(5):055002, 2018.

\bibitem{Corgier20}
Robin Corgier, Sina Loriani, Holger Ahlers, Katerine {Posso-Trujillo},
  Christian Schubert, Ernst~M. Rasel, Eric Charron, and Naceur Gaaloul.
\newblock Interacting quantum mixtures for precision atom interferometry.
\newblock {\em New Journal of Physics}, 22(12):123008, December 2020.

\bibitem{Deppner2021}
Christian Deppner, Waldemar Herr, Merle Cornelius, Peter Stromberger, Tammo
  Sternke, Christoph Grzeschik, Alexander Grote, Jan Rudolph, Sven Herrmann,
  Markus Krutzik, et~al.
\newblock Collective-mode enhanced matter-wave optics.
\newblock {\em Phys. Rev. Lett.}, 127(10):100401, 2021.

\bibitem{Gaaloul22}
Naceur Gaaloul, Matthias Meister, Robin Corgier, Annie Pichery, Patrick Boegel,
  Waldemar Herr, Holger Ahlers, Eric Charron, Jason~R. Williams, Robert~J.
  Thompson, Wolfgang~P. Schleich, Ernst~M. Rasel, and Nicholas~P. Bigelow.
\newblock {A space-based quantum gas laboratory at picokelvin energy scales}.
\newblock {\em Nature Communications}, 13:7889, 2022.

\bibitem{Condon2019}
G.~Condon, M.~Rabault, B.~Barrett, L.~Chichet, R.~Arguel, H.~Eneriz-Imaz,
  D.~Naik, A.~Bertoldi, B.~Battelier, P.~Bouyer, and A.~Landragin.
\newblock All-optical bose-einstein condensates in microgravity.
\newblock {\em Phys. Rev. Lett.}, 123:240402, Dec 2019.

\bibitem{barrett2016}
B.~Barrett, L.~Antoni-Micollier, L.~Chichet, B.~Battelier, T.~Lévèque,
  A.~Landragin, and P.~Bouyer.
\newblock {Dual Matter-Wave Inertial Sensors in Weightlessness}.
\newblock {\em Nature Communications}, 7:2041--1723, 2016.

\bibitem{Geiger2011}
R.~Geiger, V.~Ménoret, G.~Stern, N.~Zahzam, P.~Cheinet, B.~Battelier,
  A.~Villing, F.~Moron, M.~Lours, Y.~Bidel, A.~Bresson, A.~Landragin, and
  P.~Bouyer.
\newblock {Detecting inertial effects with airborne matter-wave
  interferometry}.
\newblock {\em Nature Communications}, 2:474, 2011.

\bibitem{Rudolph2015}
Jan Rudolph, Waldemar Herr, Christoph Grzeschik, Tammo Sternke, Alexander
  Grote, Manuel Popp, Dennis Becker, Hauke Müntinga, Holger Ahlers, Achim
  Peters, Claus Lämmerzahl, Klaus Sengstock, Naceur Gaaloul, Wolfgang Ertmer,
  and Ernst~M Rasel.
\newblock A high-flux {BEC} source for mobile atom interferometers.
\newblock {\em New Journal of Physics}, 17(6):065001, June 2015.

\bibitem{vanZoest2010}
Tim van Zoest, N~Gaaloul, Y~Singh, H~Ahlers, W~Herr, ST~Seidel, W~Ertmer,
  E~Rasel, Michael Eckart, Endre Kajari, et~al.
\newblock Bose-einstein condensation in microgravity.
\newblock {\em Science}, 328(5985):1540--1543, 2010.

\bibitem{Vogt2020}
Christian Vogt, Marian Woltmann, Sven Herrmann, Claus L\"ammerzahl, Henning
  Albers, Dennis Schlippert, and Ernst~M. Rasel.
\newblock Evaporative cooling from an optical dipole trap in microgravity.
\newblock {\em Phys. Rev. A}, 101:013634, Jan 2020.

\bibitem{Kulas2017}
Sascha Kulas, Christian Vogt, Andreas Resch, Jonas Hartwig, and Sven {Ganske et
  al.}
\newblock Miniaturized lab system for future cold atom experiments in
  microgravity.
\newblock {\em Microgravity Science and Technology}, 29:37, 2017.

\bibitem{becker2018}
Dennis Becker, Maike~D. Lachmann, Stephan~T. Seidel, Holger Ahlers, Aline~N.
  Dinkelaker, Jens Grosse, Ortwin Hellmig, Hauke Müntinga, Vladimir Schkolnik,
  Thijs Wendrich, André Wenzlawski, Benjamin Weps, Robin Corgier, Tobias
  Franz, Naceur Gaaloul, Waldemar Herr, Daniel Lüdtke, Manuel Popp, Sirine
  Amri, Hannes Duncker, Maik Erbe, Anja Kohfeldt, André Kubelka-Lange, Claus
  Braxmaier, Eric Charron, Wolfgang Ertmer, Markus Krutzik, Claus Lämmerzahl,
  Achim Peters, Wolfgang~P. Schleich, Klaus Sengstock, Reinhold Walser, Andreas
  Wicht, Patrick Windpassinger, and Ernst~M. Rasel.
\newblock Space-borne {Bose}–{Einstein} condensation for precision
  interferometry.
\newblock {\em Nature}, 562(7727):391--395, October 2018.

\bibitem{Lachmann2021}
Maike~D. Lachmann, Holger Ahlers, Dennis Becker, Aline~N. Dinkelaker, Jens
  Grosse, Ortwin Hellmig, Hauke Müntinga, Vladimir Schkolnik, Stephan~T.
  Seidel, Thijs Wendrich, André Wenzlawski, Benjamin Carrick, Naceur Gaaloul,
  Daniel Lüdtke, Claus Braxmaier, Wolfgang Ertmer, Markus Krutzik, Claus
  Lämmerzahl, Achim Peters, Wolfgang~P. Schleich, Klaus Sengstock, Andreas
  Wicht, Patrick Windpassinger, and Ernst~M. Rasel.
\newblock Ultracold atom interferometry in space.
\newblock {\em Nature Communications}, 12(1317):2041--1723, February 2021.

\bibitem{aveline2020}
David~C Aveline, Jason~R Williams, Ethan~R Elliott, Chelsea Dutenhoffer,
  James~R Kellogg, James~M Kohel, Norman~E Lay, Kamal Oudrhiri, Robert~F
  Shotwell, Nan Yu, et~al.
\newblock Observation of bose--einstein condensates in an earth-orbiting
  research lab.
\newblock {\em Nature}, 582(7811):193--197, 2020.

\bibitem{JPL23}
E.~R.~Elliott et. al.
\newblock Quantum gas mixtures and dual-species atom interferometry in space.
\newblock \url{https://arxiv.org/pdf/2306.15223.pdf}, 2023.

\bibitem{levequePRL}
T.~L\'ev\`eque, A.~Gauguet, F.~Michaud, F.~Pereira Dos~Santos, and
  A.~Landragin.
\newblock Enhancing the area of a raman atom interferometer using a versatile
  double-diffraction technique.
\newblock {\em Phys. Rev. Lett.}, 103:080405, Aug 2009.

\bibitem{AhlersPRL}
H.~Ahlers, H.~M\"untinga, A.~Wenzlawski, M.~Krutzik, G.~Tackmann, S.~Abend,
  N.~Gaaloul, E.~Giese, A.~Roura, R.~Kuhl, C.~L\"ammerzahl, A.~Peters,
  P.~Windpassinger, K.~Sengstock, W.~P. Schleich, W.~Ertmer, and E.~M. Rasel.
\newblock Double bragg interferometry.
\newblock {\em Phys. Rev. Lett.}, 116:173601, Apr 2016.

\bibitem{Hartmann2020}
Sabrina Hartmann, Jens Jenewein, Enno Giese, Sven Abend, Albert Roura, Ernst~M.
  Rasel, and Wolfgang~P. Schleich.
\newblock Regimes of atomic diffraction: Raman versus bragg diffraction in
  retroreflective geometries.
\newblock {\em Phys. Rev. A}, 101:053610, May 2020.

\bibitem{loriani2020}
Sina Loriani, Christian Schubert, Dennis Schlippert, Wolfgang Ertmer,
  Franck~Pereira Dos~Santos, Ernst~Maria Rasel, Naceur Gaaloul, and Peter Wolf.
\newblock Resolution of the colocation problem in satellite quantum tests of
  the universality of free fall.
\newblock {\em Physical Review D}, 102(12):124043, 2020.

\bibitem{Ahlers2022}
Holger Ahlers, Leonardo Badurina, Angelo Bassi, Baptiste Battelier, Quentin
  Beaufils, Kai Bongs, Philippe Bouyer, Claus Braxmaier, Oliver Buchmueller,
  Matteo Carlesso, Eric Charron, Maria~Luisa Chiofalo, Robin Corgier, Sandro
  Donadi, Fabien Droz, Robert Ecoffet, John Ellis, Frédéric Estève, Naceur
  Gaaloul, Domenico Gerardi, Enno Giese, Jens Grosse, Aurélien Hees, Thomas
  Hensel, Waldemar Herr, Philippe Jetzer, Gina Kleinsteinberg, Carsten Klempt,
  Steve Lecomte, Louise Lopes, Sina Loriani, Gilles Métris, Thierry Martin,
  Victor Martín, Gabriel Müller, Miquel Nofrarias, Franck Pereira~Dos Santos,
  Ernst~M. Rasel, Alain Robert, Noah Saks, Mike Salter, Dennis Schlippert,
  Christian Schubert, Thilo Schuldt, Carlos~F. Sopuerta, Christian Struckmann,
  Guglielmo~M. Tino, Tristan Valenzuela, Wolf von Klitzing, Lisa Wörner, Peter
  Wolf, Nan Yu, and Martin Zelan.
\newblock Ste-quest: Space time explorer and quantum equivalence principle
  space test.
\newblock 2022.

\bibitem{Touboul2017}
Pierre Touboul, Gilles M{\'e}tris, Manuel Rodrigues, Yves Andr{\'e}, Quentin
  Baghi, Jo{\"e}l Berg{\'e}, Damien Boulanger, Stefanie Bremer, Patrice Carle,
  Ratana Chhun, et~al.
\newblock Microscope mission: first results of a space test of the equivalence
  principle.
\newblock {\em Physical review letters}, 119(23):231101, 2017.

\bibitem{Voyage2050}
Baptiste Battelier, Joël Bergé, Andrea Bertoldi, Luc Blanchet, Kai Bongs,
  Philippe Bouyer, Claus Braxmaier, Davide Calonico, Pierre Fayet, Naceur
  Gaaloul, Christine Guerlin, Aurélien Hees, Philippe Jetzer, Claus
  Lämmerzahl, Steve Lecomte, Christophe~Le Poncin-Lafitte, Sina Loriani,
  Gilles Métris, Miquel Nofrarias, Ernst Rasel, Serge Reynaud, Manuel
  Rodrigues, Markus Rothacher, Albert Roura, Christophe Salomon, Stephan
  Schiller, Wolfgang~P. Schleich, Christian Schubert, Carlos~F. Sopuerta,
  Fiodor Sorrentino, Timothy~J. Sumner, Guglielmo~M. Tino, Philip Tuckey, Wolf
  von Klitzing, Lisa Wörner, Peter Wolf, and Martin Zelan.
\newblock Exploring the foundations of the physical universe with space tests
  of the equivalence principle.
\newblock {\em Exp Astron}, 51:1695–1736, 2021.

\bibitem{Overstreet2022}
Chris Overstreet, Peter Asenbaum, Joseph Curti, Minjeong Kim, and Mark~A.
  Kasevich.
\newblock Observation of a gravitational aharonov-bohm effect.
\newblock {\em Science}, 375(6577):226--229, 2022.

\bibitem{ufrecht2020}
Christian Ufrecht and Enno Giese.
\newblock Perturbative operator approach to high-precision light-pulse atom
  interferometry.
\newblock {\em Phys. Rev. A}, 101:053615, May 2020.

\bibitem{Overstreet2021}
Chris Overstreet, Peter Asenbaum, and Mark~A. Kasevich.
\newblock Physically significant phase shifts in matter-wave interferometry.
\newblock {\em American Journal of Physics}, 89(3):324--332, 2021.

\bibitem{Roura2017a}
Albert Roura.
\newblock Circumventing heisenberg’s uncertainty principle in atom
  interferometry tests of the equivalence principle.
\newblock {\em Physical Review Letters}, 118:160401, 2017.

\bibitem{Haslinger2018}
Philipp Haslinger, Matt Jaffe, Victoria Xu, Osip Schwartz, Matthias
  Sonnleitner, Monika Ritsch-Marte, Helmut Ritsch, and Holger Müller.
\newblock Attractive force on atoms due to blackbody radiation.
\newblock {\em Nature Physics}, 14(3):257--260, 2018.

\bibitem{Vanier1989}
J.~Vanier and C.~Audoin.
\newblock {\em The Quantum Physics of Atomic Frequency Standards: Volume I}.
\newblock Adam Hilger, 1989.

\bibitem{Wolf2006a}
P.~Wolf, F.~Chapelet, S.~Bize, and A.~Clairon.
\newblock Cold atom clock test of lorentz invariance in the matter sector.
\newblock {\em Phys. Rev. Lett.}, 96:060801, 2006.

\bibitem{Wodey2020}
E.~Wodey, D.~Tell, E.~M. Rasel, D.~Schlippert, R.~Baur, U.~Kissling,
  B.~Kölliker, M.~Lorenz, M.~Marrer, U.~Schläpfer, M.~Widmer, C.~Ufrecht,
  S.~Stuiber, and P.~Fierlinger.
\newblock {A scalable high-performance magnetic shield for very long baseline
  atom interferometry}.
\newblock {\em Review of Scientific Instruments}, 91(3):035117, 03 2020.

\bibitem{cheinet2008}
Patrick Cheinet, Benjamin Canuel, Franck~Pereira Dos~Santos, Alexandre Gauguet,
  Florence Yver-Leduc, and Arnaud Landragin.
\newblock Measurement of the sensitivity function in a time-domain atomic
  interferometer.
\newblock {\em IEEE Transactions on instrumentation and measurement},
  57(6):1141--1148, 2008.

\bibitem{schubert2019scalable}
C~Schubert, D~Schlippert, S~Abend, E~Giese, A~Roura, WP~Schleich, W~Ertmer, and
  EM~Rasel.
\newblock Scalable, symmetric atom interferometer for infrasound gravitational
  wave detection.
\newblock {\em arXiv preprint arXiv:1909.01951}, 2019.

\bibitem{robert2022microscope}
Alain Robert, Valerio Cipolla, Pascal Prieur, Pierre Touboul, Gilles
  M{\'e}tris, Manuel Rodrigues, Yves Andr{\'e}, Joel Berg{\'e}, Damien
  Boulanger, Ratana Chhun, et~al.
\newblock Microscope satellite and its drag-free and attitude control system.
\newblock {\em Classical and Quantum Gravity}, 39(20):204003, 2022.

\bibitem{hensel2021}
Thomas Hensel, Sina Loriani, Christian Schubert, Florian Fitzek, Sven Abend,
  Holger Ahlers, J-N Siem{\ss}, Klemens Hammerer, Ernst~Maria Rasel, and Naceur
  Gaaloul.
\newblock Inertial sensing with quantum gases: a comparative performance study
  of condensed versus thermal sources for atom interferometry.
\newblock {\em The European Physical Journal D}, 75(3):1--13, 2021.

\bibitem{duchayne2009}
Lo{\"\i}c Duchayne, Flavien Mercier, and Peter Wolf.
\newblock Orbit determination for next generation space clocks.
\newblock {\em Astronomy \& Astrophysics}, 504(2):653--661, 2009.

\bibitem{christophe2015}
Bruno Christophe, Damien Boulanger, Bernard Foulon, P-A Huynh, Vincent Lebat,
  Francoise Liorzou, and E~Perrot.
\newblock A new generation of ultra-sensitive electrostatic accelerometers for
  grace follow-on and towards the next generation gravity missions.
\newblock {\em Acta Astronautica}, 117:1--7, 2015.

\bibitem{pihan2019new}
H{\'e}l{\`e}ne Pihan-Le~Bars, Christine Guerlin, Aurelien Hees, Romain
  Peaucelle, Jay~D Tasson, Quentin~G Bailey, Geoffrey Mo, Pac{\^o}me Delva,
  Fr{\'e}d{\'e}ric Meynadier, Pierre Touboul, et~al.
\newblock New test of lorentz invariance using the microscope space mission.
\newblock {\em Physical Review Letters}, 123(23):231102, 2019.

\bibitem{armano2019}
Michele Armano, Heather Audley, Jonathon Baird, Pierre Binetruy, Michael Born,
  Daniele Bortoluzzi, Eleanora Castelli, Antonella Cavalleri, Andrea Cesarini,
  AM~Cruise, et~al.
\newblock Lisa pathfinder platform stability and drag-free performance.
\newblock {\em Physical Review D}, 99(8):082001, 2019.

\bibitem{Storey1994}
P.~Storey and C.~Cohen-Tannoudji.
\newblock The {Feynman} path integral approach to atomic interferometry. {A}
  tutorial.
\newblock {\em J. Phys. II France}, 4(11):1999--2027, November 1994.

\bibitem{Geiger18}
Remi Geiger and Michael Trupke.
\newblock Proposal for a quantum test of the weak equivalence principle with
  entangled atomic species.
\newblock {\em Phys. Rev. Lett.}, 120:043602, Jan 2018.

\bibitem{Szigeti20}
Stuart~S. Szigeti, Samuel~P. Nolan, John~D. Close, and Simon~A. Haine.
\newblock High-precision quantum-enhanced gravimetry with a bose-einstein
  condensate.
\newblock {\em Phys. Rev. Lett.}, 125:100402, Sep 2020.

\bibitem{Corgier21}
Robin Corgier, Naceur Gaaloul, Augusto Smerzi, and Luca Pezz\`e.
\newblock Delta-kick squeezing.
\newblock {\em Phys. Rev. Lett.}, 127:183401, Oct 2021.

\bibitem{Corgier23}
Robin Corgier, Marco Malitesta, Augusto Smerzi, and Luca Pezz{\`{e}}.
\newblock Quantum-enhanced differential atom interferometers and clocks with
  spin-squeezing swapping.
\newblock {\em {Quantum}}, 7:965, March 2023.

\bibitem{Giese2014}
E.~Giese, S.~Kleinert, M.~Meister, V.~Tamma, A.~Roura, and W.~P. Schleich.
\newblock The interface of gravity and quantum mechanics illuminated by
  {W}igner phase space.
\newblock In G.~M. Tino and M.~A. Kasevich, editors, {\em Atom Interferometry},
  Proceedings of the International School of Physics ``Enrico Fermi'', Course
  188, page 411. IOS Press, Amsterdam, 2014.

\bibitem{Dubetsky2016}
B.~Dubetsky, S.~B. Libby, and P.~Berman.
\newblock Atom interferometry in the presence of an external test mass.
\newblock {\em Atoms}, 4(2), 2016.

\bibitem{Antoine2003}
C.~Antoine and C.J. Bord\'e.
\newblock Exact phase shifts for atom interferometry.
\newblock {\em Phys. Lett. A}, 306(5):277--284, 2003.

\bibitem{Schleich2013}
W.~P. Schleich, D.~M. Greenberger, and E.~M. Rasel.
\newblock Redshift controversy in atom interferometry: {Representation}
  dependence of the origin of phase shift.
\newblock {\em Phys. Rev. Lett.}, 110(1):010401, January 2013.

\bibitem{Kleinert2015}
Stephan Kleinert, Endre Kajari, Albert Roura, and Wolfgang~P. Schleich.
\newblock Representation-free description of light-pulse atom interferometry
  including non-inertial effects.
\newblock {\em Phys. Rep.}, 605:1 -- 50, 2015.

\bibitem{Feynman1965}
R.~P. Feynman and A.~P. Hibbs.
\newblock {\em Quantum Mechanics and Path Integrals}.
\newblock McGraw-Hill, New York, 1965.

\bibitem{Su1988}
Z.-b. Su, L.-Y. Chen, X.-t. Yu, and K.-c. Chou.
\newblock Influence functional and closed-time-path {G}reen's function.
\newblock {\em Phys. Rev. B}, 37:9810--9812, Jun 1988.

\bibitem{DeWitt2003}
B.~S. DeWitt.
\newblock {\em The global approach to quantum field theory}, volume~1.
\newblock Oxford University Press, Oxford, 2003.

\bibitem{Feldbrugge2023}
J.~Feldbrugge and N.~Turok.
\newblock Existence of real time quantum path integrals.
\newblock {\em Ann. Phys.}, 454:169315, 2023.

\bibitem{Roura2014}
A.~Roura, W.~Zeller, and W.~P. Schleich.
\newblock Overcoming loss of contrast in atom interferometry due to gravity
  gradients.
\newblock {\em New J. Phys.}, 16(12):123012, dec 2014.

\bibitem{Loriani2019}
S.~Loriani, A.~Friedrich, C.~Ufrecht, F.~Di~Pumpo, S.~Kleinert, S.~Abend,
  N.~Gaaloul, C.~Meiners, C.~Schubert, D.~Tell, {\'{E}}.~Wodey, M.~Zych,
  W.~Ertmer, A.~Roura, D.~Schlippert, W.~P. Schleich, E.~M. Rasel, and
  E.~Giese.
\newblock Interference of clocks: {A} quantum twin paradox.
\newblock {\em Sci. Adv.}, 5(10):eaax8966, October 2019.

\end{thebibliography}

\end{document}